\documentclass[reprint,amsmath,amssymb,aps,nofootinbib]{revtex4-2}

\newcommand{\warp}{\textsc{warp}}

\usepackage[T1]{fontenc}
\usepackage{graphicx}
\usepackage{dcolumn}
\usepackage{bm}
\usepackage{booktabs}
\usepackage{tabularx}
\usepackage[mathlines]{lineno}
\usepackage{mathtools}
\usepackage[colorlinks,linkcolor=blue,citecolor=blue,urlcolor=blue]{hyperref}
\usepackage{mathrsfs}
\usepackage{float}
\usepackage{subcaption}
\usepackage{tikz}
\usetikzlibrary{arrows.meta,positioning,shapes.geometric,fit,backgrounds,calc}

\usepackage{ragged2e}
\makeatletter
\let\SIsave@maketitle\maketitle\let\SIsave@title\title\let\SIsave@author\author
\let\SIsave@email\email\let\SIsave@affiliation\affiliation\let\SIsave@date\date
\makeatother
\begin{document}

\title{Finite temperature and pressure structural relaxations including zero-point motion and anharmonicity}

\author{Henricus R.\,A. ten Eikelder$^{1}$}
\email{henricus.ten.eikelder@liu.se}
\author{Florian Knoop$^{1}$}
\author{Florian Trybel$^{1}$}
\email{florian.trybel@liu.se}
\affiliation{1) Department of Physics, Chemistry and Biology (IFM), Link\"oping University, SE-581 83 Link\"oping, Sweden}

\date{\today}

\begin{abstract}
At finite temperature and pressure, the equilibrium structure of a crystal minimizes the Gibbs free energy rather than the static Born–Oppenheimer energy. We present \warp{} (Wyckoff-constrained Anharmonic Relaxations at Pressure and temperature), a symmetry-constrained, variable-cell relaxation based on the stochastic temperature-dependent effective potential method, including anharmonicity and zero-point motion. Every position update is projected onto Wyckoff positions and every strain onto symmetric strains, so a relaxation can raise the symmetry but never lower it. For aluminum, the thermal expansion follows measurements at cryogenic temperatures, including the quantum plateau below 50~K. For lanthanum hydrides at megabar pressures, equations of state agree with phonon-based and path-integral calculations, and LaH$_{10}$ relaxes from its rhombohedral static minimum to the observed cubic phase, also with a machine-learned potential fine-tuned on 150 supercells. For brucite, Mg(OD)$_2$, lattice parameters, bond distances and thermal expansion agree with neutron diffraction, and the O–D isotope effect is captured.
\end{abstract}
\maketitle

\section{Introduction}

At finite temperature and pressure the equilibrium structure of a crystal minimizes its Gibbs free energy rather than its static enthalpy \cite{BornHuang2002}. This is well documented across several material classes: hydrogen-rich compounds, a promising class of materials in the search for high-temperature superconductivity \cite{errea2020quantum,laniel2022high,PIMD_Ackland,hydrogen-q-3,errea_2025}, hydrous minerals relevant for the structure and dynamics in planetary interiors \cite{hermann2016brucite,schaack2020brucite,sanofurukawa2018alooh,bronstein2017alooh}, perovskites relevant to optoelectronics \cite{klarbring2020anharmonicity}, and thermoelectric materials \cite{ribeiro2018strong}. 

Hydrogen is the element whose position is hardest to measure, in particular at high pressure. Hydrogen positions can in general not be determined from X-ray diffraction, and neutron diffraction is not commonly available above 25~GPa \cite{haberl_advancing_2023,hydrogen-pickard-nmr}. Therefore, hydrogen content and positions are routinely inferred by matching the heavy-atom sublattice and volume to calculation. Such a comparison is only as good as the underlying calculation, where often temperature is not taken into account, which can have a significant influence on the actual finite temperature volume.

Two established routes to include temperature and nuclear quantum effects exist. Path-integral molecular dynamics (PIMD) treats the nuclei quantum mechanically \cite{hydrogen-pimd,Hydrogen_pimd2,PIMD-ML_1,PIMD_Ackland} and, with machine-learned interatomic potentials (MLIPs), has become computationally feasible even for large systems \cite{PIMD_ML3}; but it remains demanding. The extraction of high-symmetry equilibrium positions from a trajectory is not straightforward, and the reliability of pre-trained potentials under extreme conditions is an open question \cite{Loew2026}. The second route is based on effective, finite temperature phonons. As for example in the stochastic self-consistent harmonic approximation (SSCHA) \cite{sscha,gao2025iterativelearningschemecrystal}, the self-consistent phonon (SCPH) method as implemented in Alamode \cite{alamode,tadano2015scph,masuki2022scph} and the temperature-dependent effective potential (TDEP) method \cite{TDEP_1st,TDEP_opensourcesoftware,dubrovinsky2022materials,TDEP_Si,laniel2022high,TDEPfk}; hiPhive \cite{hiphive}, by contrast, is a code for fitting force constants rather than a method. SCPH and TDEP fit force constants of arbitrary order to displacement--force data and calculate renormalised phonons and the properties derived from them; the reference structure around which the force constants are expanded is an input to the fit rather than an output of it. SSCHA allows relaxation of positions and cell on the free-energy surface \cite{sscha,errea2020quantum}, symmetrising the structure, the gradient and the force constants at every step. Machine-learned interatomic potentials can reduce the cost of path-integral simulations \cite{PIMD_ML3} and SSCHA calculations \cite{belli2025efficient,dong2025machine}, while foundation-model-assisted workflows have also been developed for anharmonic crystal structure prediction \cite{gao2025iterativelearningschemecrystal,poletaev2025sscha}. Finite-temperature structural optimization is already available within SSCHA \cite{sscha,monacelli2018stress} and self-consistent phonon theory \cite{masuki2022scph}, while TDEP includes residual-force-based optimization of internal coordinates \cite{TDEP_2nd}. 

\begin{figure*}[t]
    \centering
    \includegraphics[width=1.11\linewidth]{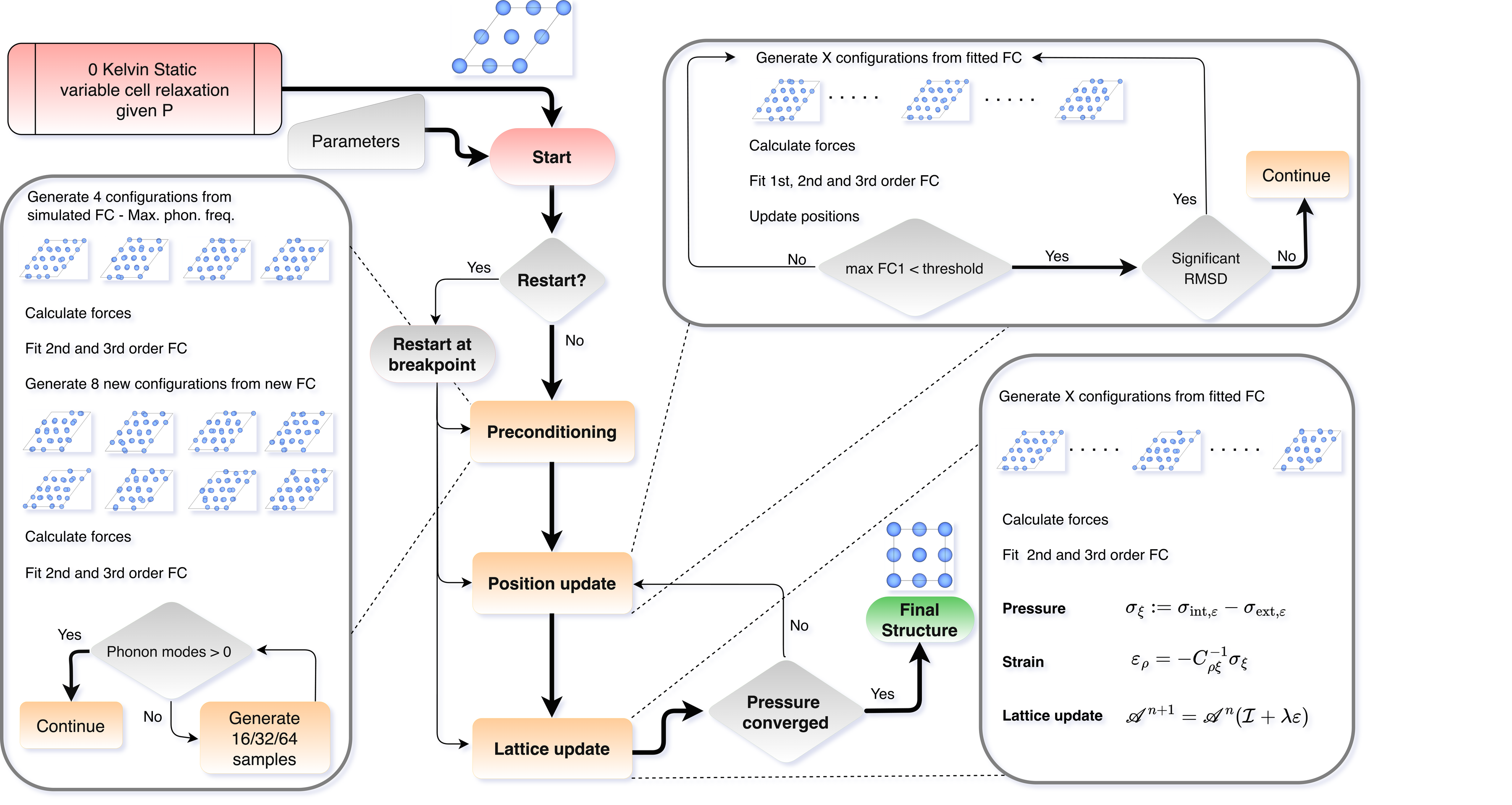}
      \caption{\justifying \textbf{The \warp{} relaxation loop.} A preconditioning stage builds a stable set of model force constants; the position loop then samples thermally displaced supercells, fits force constants, and moves the atoms by the Wyckoff-projected first-order displacement until the residual first-order force constants fall below tolerance; the lattice step subtracts the target stress from the internal stress of Eq.~\eqref{eq:sigma_int} and moves the cell by the symmetry-projected strain. After a lattice step the previous ensemble is no longer representative and is discarded. The loop closes when the internal stress matches the target within tolerance. The position-loop box lists the orders that are fitted there and not the sequence in which they are fitted, which is the second order, then the first-order constants $\Phi^{(1)}$ from the residual of Eq.~\eqref{eq:fc1}, then the third order fitted to what is left. The ``max FC1'' of that box is $\max_i|\Phi^{(1)}_i|$; the convergence gate of Sec.~\ref{sec:workflow} is on the mean as well as the maximum.}
    \label{fig:flowchart}
\end{figure*}

The method developed in this work, \warp{} (Wyckoff-constrained Anharmonic Relaxations at Pressure and temperature), built on stochastic TDEP (sTDEP) \cite{TDEP_s-tdep}. It relaxes the internal coordinates and lattice parameters at a given temperature and target stress tensor, including anharmonicity and zero-point motion. The relaxation workflow is shown in Fig.~\ref{fig:flowchart}. Every position update is projected onto the Wyckoff degrees of freedom of the current space group, and every cell update onto the symmetric strains of its point group. The relaxation is therefore symmetry-constrained by construction: it can raise the symmetry of a structure but never lower it, and the symmetry-forbidden part of the sampling noise never reaches the structure. Forces can come from density functional theory (DFT) or from a machine-learned interatomic potential through an ASE interface \cite{larsen2017ase}. Sensible defaults are automatically chosen, so beyond a few physically meaningful tolerances and the sampling settings listed per system in Supplementary Section~S3 the method needs no tuning. In this paper  \warp{} is demonstrated on three test cases of increasing complexity: (i)~the thermal expansion of aluminium down to cryogenic temperatures, which tests the quantum statistics of the sampling; (ii)~the equations of state of four lanthanum hydrides at megabar pressures, together with the sub- to supergroup transition of LaH$_{10}$ previously found at finite temperature ~\cite{errea2020quantum}; and (iii)~brucite, Mg(OD)$_2$, a layered hydrous mineral whose hydrogen positions have been measured directly by neutron diffraction from 15 to 300~K, against which the lattice parameters, the layer geometry and the O--D distance are compared. Throughout, supercells obtained during the DFT-based sampling double as training sets for the fine-tuning of machine learned interatomic potentials, which allows further decreasing the computational cost of \warp{} relaxations by orders of magnitude.The output of \warp{} is a structure at which the thermally averaged forces and stress vanish.

\section{Methods}
The subsections up to Sec.~\ref{sec:workflow} describe the method, i.e.\ the algorithmic steps of initial guess, sampling, force-constant fitting, symmetrisation and update of positions and cell; Sec.~\ref{sec:mlip} and Supplementary Sections~S3--S4 describe the implementation, i.e.\ the force engines behind the ASE interface, the defaults and the settings used per system.

\subsection{The effective Hamiltonian and stochastic sampling}
The TDEP method fits the force constants of a model Hamiltonian $H=\sum_i\mathbf{p}_i^2/2m_i+U$ \cite{TDEP_1st,TDEP_2nd,TDEP_fc3}, whose potential energy is a model potential in the displacements $u$ of a reference structure $\mathbf{R}^{0}$. U has the form,
\begin{equation}
\begin{split}
 U = U_0 &+ \sum_{i\alpha}\Phi^{(1),\alpha}_{i}u^{\alpha}_i \\
&+ \frac{1}{2!}\sum_{ij\alpha\beta}\Phi^{(2),\alpha\beta}_{ij}u^\alpha_iu^\beta_j \\
&+ \frac{1}{3!} \sum_{ijk\alpha\beta\gamma}\Phi^{(3),\alpha\beta\gamma}_{ijk}u^\alpha_iu^\beta_ju^\gamma_k,
\end{split}
\label{hamiltonian}
\end{equation}
and is fitted to forces computed on displaced supercells. $U_0$ is the fitted baseline energy of Eq.~(21) of \citet{TDEP_2nd}: the mean sampled potential energy minus the second-order model energy, written $U_0^{\mathrm{TDEP}}$ in Eq.~\eqref{eq:U0} below; $u$ are the displacements, roman indices label atoms and greek indices Cartesian directions, and $\Phi^{(1)}$, $\Phi^{(2)}$ and $\Phi^{(3)}$ are the first-, second- and third-order force constants; the superscript in parentheses gives the order. These are explicit functions of the reference positions and of temperature, hence the name of the method: $\Phi^{(1)}(\mathbf{R}^{0},T)$, $\Phi^{(2)}(\mathbf{R}^{0},T)$ and $\Phi^{(3)}(\mathbf{R}^{0},T)$ are effective force constants of the model at the current reference of the iterative process. They are obtained from the fit over the sampled ensemble, and not derivatives of the Born--Oppenheimer surface. In sTDEP \cite{TDEP_s-tdep}, which is used throughout this work, the displaced supercells are drawn from a harmonic canonical ensemble built from the second-order force constants alone; $\Phi^{(1)}$ and $\Phi^{(3)}$ are fitted to the residual forces \cite{castellano2023} and do not enter the sampling: $\Phi^{(1)}$ enters only the position update (Sec.~\ref{sec:posupdate}) and $\Phi^{(3)}$ only the stress of the lattice update, Eq.~(\ref{eq:sigma_int}). In this ensemble the displacements $u_{i}$ are described by,
\begin{equation}
u_i = \sum_{s=1}^{3N_a}\varepsilon_{is} \left< A_{is}\right> \sqrt{-2\ln\xi_{1s}}\,\sin 2\pi\xi_{2s},
\label{eq:sampling}
\end{equation}
\begin{equation}
\left< A_{is} \right> = \sqrt{\frac{\hbar(2n_s +1 )}{2m_i\omega_s}}
\;\;\xrightarrow[\;\hbar\omega_s\ll k_BT\;]{}\;\;
\frac{1}{\omega_s}\sqrt{\frac{k_BT}{m_i}},
\label{eq:amplitude}
\end{equation}
where $N_a$ is the number of atoms in the supercell, $\omega_s$ the frequency of mode $s$ of the current second-order force constants, $\varepsilon_{is}$ its eigenvector on atom $i$, $n_s=[\exp(\hbar\omega_s/(k_BT))-1]^{-1}$ is the Bose--Einstein occupation. $\xi_{1s}$, $\xi_{2s}$ are uniform random numbers on $(0,1)$, drawn independently for every mode of every configuration. The combination $\sqrt{-2\ln\xi_{1s}}\sin(2\pi\xi_{2s})$ is a Box--Muller transform, so every mode enters with an independent Gaussian amplitude, whose root-mean-square value on atom $i$ is $\langle A_{is}\rangle$. Given that $2n_s+1=\coth(\hbar\omega_s/(2k_BT))$, the square of $\langle A_{is}\rangle$ is then given by $\hbar\coth(\hbar\omega_s/(2k_BT))/(2m_i\omega_s)$ which is the exact thermal mean-square amplitude of a quantized harmonic oscillator in the canonical ensemble \cite{TDEP_s-tdep,west2006}. The displacements, as sums of independent Gaussians, follow the thermal distribution of the harmonic crystal. Every configuration is therefore an uncorrelated draw, and its displacements include the zero-point amplitude from the first step.

Equation~\eqref{eq:amplitude} reduces to the two limits that bracket the problem. For $\hbar\omega_s\gg k_BT$ the occupation $n_s$ vanishes, $2n_s+1\to1$, and the amplitude becomes $\sqrt{\hbar/2m_i\omega_s}$: the zero-point amplitude of the harmonic ground state, which is finite at $T=0$. For $\hbar\omega_s\ll k_BT$ the hyperbolic cotangent approaches $2k_BT/\hbar\omega_s$, the $\hbar$ cancels, and the amplitude reduces to $\omega_s^{-1}\sqrt{k_BT/m_i}$---the equipartition result, independent of Planck's constant.

In the presented examples, quantum-mechanical sampling is used, the cost is identical, since the two distributions differ only in the width assigned to each displacement. A mode with negative squared frequency receives zero amplitude in the sampled configurations (TDEP assigns no displacement to non-positive frequencies). 

The stochastic sampling needs starting force constants to generate the first displacements. These can be harmonic force constants from a finite-displacement calculation, for instance with \textsc{Phonopy}~\cite{togo2015phonopy}, or the model force constants of \citet{TDEP_s-tdep}, which are generated from a single parameter, the Debye temperature or an estimate of the maximum phonon frequency, taken for instance from experimental vibrational data. The preconditioning stage described below then replaces them with fitted ones.

For any Hamiltonian the derivative of the canonical free energy with respect to a parameter is the thermal average of the derivative of the Hamiltonian, $\partial F/\partial\lambda=\langle\partial H/\partial\lambda\rangle_T$. The equilibrium structure at temperature $T$ and target stress is therefore the one at which the thermally averaged forces vanish and the thermally averaged stress, nuclear kinetic contribution included, equals the target \cite{monacelli2018stress}; \warp{} evaluates these two averages over the sTDEP ensemble (Sec.~\ref{sec:posupdate} and the lattice update below) and moves the structure until they hold. The harmonic free energy of the converged model, $F = U_0^{\mathrm{TDEP}} + F_{\mathrm{ph}}$ with
\begin{equation}
F_{\mathrm{ph}} = \sum_{\mathbf{q}s}\Bigl[\tfrac12\hbar\omega_{\mathbf{q}s} + k_BT\ln\bigl(1-e^{-\beta\hbar\omega_{\mathbf{q}s}}\bigr)\Bigr]
\end{equation}
and
\begin{equation}
U_0^{\mathrm{TDEP}} = \Bigl\langle U - \tfrac{1}{2!}\sum_{ij\alpha\beta}\Phi^{(2),\alpha\beta}_{ij}u^{\alpha}_iu^{\beta}_j \Bigr\rangle ,
\label{eq:U0}
\end{equation}
the mean sampled potential energy minus the second-order model energy of the samples, is the Gibbs--Bogoliubov bound whose stationary point the self-consistent sampling approaches \cite{TDEP_2nd,sscha}.

\subsection{Position update and Wyckoff projection}
\label{sec:posupdate}
The forces on the sampled configurations are fitted to the expansion of Eq.~\eqref{hamiltonian}. If the reference structure is not at the finite-temperature equilibrium, these residual forces contain a systematic component. The orders are fitted in sequence \cite{castellano2023}, i.e. the first-order force constants are fitted to the residual after the second order and before the third,
\begin{equation}
\Phi^{(1),\alpha}_{i} = -\bigl\langle f^{\mathrm{res},(2),\alpha}_i \bigr\rangle = -\bigl\langle f^{\alpha}_i - f^{(2),\alpha}_i \bigr\rangle ,
\label{eq:fc1}
\end{equation}
where $i$ labels the atom and $\alpha$ the Cartesian direction. A non-zero $\Phi^{(1)}$ means there is a force acting on the atoms. Within the harmonic model built from the fitted second-order constants this force is
\begin{equation}
f^{\alpha}_i = -\sum_{j\beta}\Phi^{(2),\alpha\beta}_{ij}u^{\beta}_j - \Phi^{(1),\alpha}_i .
\label{eq:harmforce}
\end{equation}
The equilibrium configuration is the one at which the forces vanish, $f^{\alpha}_i=0$. Because the second-order force-constant matrix is singular by translational invariance, Eq.~\eqref{eq:harmforce} is solved for the displacement with the Moore--Penrose pseudo-inverse $(\Phi^{(2)})^{+}$, which gives the least-squares solution
\begin{equation}
(u_i^\alpha)^{\mathrm{eq}} = - \sum_{j\beta} \bigl((\Phi^{(2)})^{+}\bigr)^{\alpha\beta}_{ij}\,\Phi_j^{(1),\beta} ,
\label{eq:ueq}
\end{equation}
and the resulting equilibrium positions with applied displacements are
\begin{equation}
\mathbf{R}^{\mathrm{new}} = \mathbf{R}^{0} + \mathbf{u}^{\mathrm{eq}} .
\label{eq:rnew}
\end{equation}

The force constants themselves are symmetry-reduced in the fit, so $\Phi^{(1)}$ obeys the space-group symmetries. However, the displacement field is a least-squares solution built from a finite number of sampled forces, and its statistical noise is not symmetric. Applied repeatedly, this noise can lead to an accumulation of error and consequently a decrease in symmetry. The update is therefore projected onto the Wyckoff degrees of freedom before it is applied. At every iteration the space group of the current cell is determined
with \textsc{spglib} \cite{togo2018spglib}, and the atoms are grouped into Wyckoff orbits, sets of atoms related to each other by the symmetry operations $(\mathbf{Q}_k,\mathbf{t}_k)$ of the group, with rotation $\mathbf{Q}_k$ and translation $\mathbf{t}_k$. The site symmetry of each representative leaves between zero and three degrees of freedom. A site $(x,x,0)$ can move only along $(1,1,0)$, a site $(0,0,z)$ only along $z$, and a site $(\tfrac14,\tfrac14,\tfrac14)$ not at all. Collecting these allowed directions in a matrix $\mathsf{A}^{(a)}$ for orbit $a$, a symmetry-compatible displacement of the orbit is 
\begin{equation}
\mathbf{u}_k^{(a)} = \mathbf{Q}_k\,\mathsf{A}^{(a)}\,\mathbf{w}^{(a)},\qquad k=1,\dots,m_a,
\label{eq:wyckoff_model}
\end{equation}
with one parameter vector $\mathbf{w}^{(a)}$ per orbit, in the fractional coordinates of the cell; translations cancel because displacements are differences of positions. Given the raw displacements $\tilde{\mathbf{u}}_k^{(a)}$, $\mathbf{w}^{(a)}$ is the least-squares solution over the whole orbit,
\begin{equation}
\mathbf{w}^{(a)} = \arg\min_{\mathbf{w}}\sum_{k=1}^{m_a}\bigl\lVert \tilde{\mathbf{u}}_k^{(a)} - \mathbf{Q}_k\mathsf{A}^{(a)}\mathbf{w}\bigr\rVert^2,
\label{eq:wyckoff_lsq}
\end{equation}
an overdetermined system since $n_a\le3\le3m_a$. The update is then made in parameter space: the site parameters of the orbit are moved by $\lambda_w\,\mathbf{w}^{(a)}$, where $\lambda_w$ is a damping factor on the position update; 0.8 in this work unless stated otherwise and left to the user. Every member is regenerated from the new parameters by its symmetry operation, so the orbit stays to numerical tolerance on its Wyckoff position. Orbits with $n_a=0$ receive no displacement. The residual $\varrho_a=[(3m_a)^{-1}\sum_k\lVert\tilde{\mathbf{u}}_k^{(a)}-\mathbf{u}_k^{(a)}\rVert^2]^{1/2}$ is the symmetry-forbidden part of the raw field: a direct measure of the sampling noise, and a check that the ensemble is large enough.

Three properties follow: (i)~the projection maps onto the invariant subspace of the space group, so the update cannot lower the symmetry, by construction rather than by tolerance; (ii)~it can raise it, since the higher-symmetry structure of a supergroup is a point in the parameter space of the subgroup, reached when the free parameters take special values such as $\tfrac14$ or $\tfrac12$; and (iii)~because each parameter is averaged over all members of its orbit, its statistical error is never larger, and generally smaller, than that of a single atom's displacement. 

\subsection{Lattice update}
In the following, the convention is that stress is tension-positive and pressure compression-positive is used, i.e the pressure is minus one third of the trace of the stress. Within the model, the stress tensor can be defined as
\begin{equation}
\sigma_{\mathrm{int}} = \langle\sigma\rangle + \sigma_{\mathrm{kin}} + \delta,\qquad \delta = \sigma_{\mathrm{ph}}-\sigma_{\mathrm{FC3}},
\label{eq:sigma_int}
\end{equation}
where $\langle\sigma\rangle$ is the mean Born--Oppenheimer stress of the sampled configurations, $\sigma_{\mathrm{kin}}$ the kinetic stress of the second-order model, evaluated as the harmonic virial of the fitted second-order force constants on the same displacements, Eq.~\eqref{eq:sigma_kin}, and $\delta$ a finite-grid correction from the third-order force constants, defined below.

The first two terms carry the thermal stress. The sampled configurations are displaced on the full potential-energy surface, so $\langle\sigma\rangle$ contains the entire stress of the displacements, harmonic and anharmonic. What it lacks is the contribution of the nuclear momenta, since each configuration is a static snapshot; $\sigma_{\mathrm{kin}}$ supplies it. For the second-order model this kinetic stress is known in closed form and, by the virial theorem of the harmonic oscillator, equals the harmonic virial of the fitted force constants on the sampled displacements \cite{TDEPfk},
\begin{equation}
\begin{split}
\sigma_{\mathrm{kin}}^{\alpha\beta} &= -\frac{1}{V}\sum_{s}\hbar\omega_s\bigl(n_s+\tfrac12\bigr)\sum_i\varepsilon^{\alpha}_{is}\varepsilon^{\beta}_{is} \\
&= -\frac{1}{V}\Bigl\langle\sum_{ij\gamma}\Phi^{(2),\alpha\gamma}_{ij}\,u^{\gamma}_j u^{\beta}_i\Bigr\rangle ,
\end{split}
\label{eq:sigma_kin}
\end{equation}
with the sum over the modes $s$ of the supercell; the second form is the one used, because it cancels the harmonic part of the sampling fluctuation in $\langle\sigma\rangle$ term by term, and minus one third of its trace reduces to the ideal-gas pressure $N_a k_B T/V$ in the classical limit. For a harmonic crystal, $\sigma_{\mathrm{kin}}$ cancels the tensile virial of the displacements in $\langle\sigma\rangle$ exactly, so $\langle\sigma\rangle+\sigma_{\mathrm{kin}}$ returns the static stress and the crystal does not expand. Any departure of $\langle\sigma\rangle+\sigma_{\mathrm{kin}}$ from the static stress is therefore anharmonic: it is the thermal stress that contributes to changes in the lattice.
There are multiple ways to calculate the third order contributions to the stress. The third-order force constants carry the strain dependence of the second-order ones \cite{wallace1972,TDEPfk}, $\partial\Phi^{(2),\gamma\delta}_{ij}/\partial\varepsilon_{\alpha\beta}=\sum_k\Phi^{(3),\gamma\delta\alpha}_{ijk}R^{\beta}_k$, and with it the mode-Gr\"uneisen parameters and the Gr\"uneisen stress of the model. The correction $\delta=\sigma_{\mathrm{ph}}-\sigma_{\mathrm{FC3}}$ contains this mode-Gr\"uneisen stress of the fitted third-order force constants twice, with opposite signs. The subtracted term, $\sigma_{\mathrm{FC3}}$, evaluates it in real space on the sampled supercell; the added term, $\sigma_{\mathrm{ph}}$, evaluates it in reciprocal space which can be described as follows,
\begin{equation}
\sigma_{\mathrm{ph}} = -\frac{1}{V}\sum_{\mathbf{q}s}\gamma_{\mathbf{q}s}\,\hbar\omega_{\mathbf{q}s}\bigl(n_{\mathbf{q}s}+\tfrac12\bigr),
\label{eq:sigma_ph}
\end{equation}
where the mode Gr\"uneisen parameters, $\gamma_{\mathbf{q}s}$, come from the third-order force constants and the Bose--Einstein occupation  is given as $n_{\mathbf{q}s}$ . They differ because $\sigma_{\mathrm{FC3}}$ only contains contributions from the wavevectors commensurate with the supercell, whereas $\sigma_{\mathrm{ph}}$ can be evaluated on a denser mesh by Fourier interpolation of the same force constants. The correction $\delta$ thus replaces the supercell estimate of the Gr\"uneisen stress by its value on the denser mesh. In the limit of a large supercell and many samples the two third-order terms cancel and $\sigma_{\mathrm{kin}}$ restores the kinetic contribution that the static snapshots lack, so that $\sigma_{\mathrm{int}}$ reduces to the full thermally averaged stress of the sampled ensemble \cite{TDEPfk}; a relaxation with the second order alone converges to the same cell in that limit.

The cell is updated with the isothermal elastic constants $C$, which TDEP computes in Voigt notation from the fitted second-order force constants \cite{TDEP_opensourcesoftware}. Alternatively, a full elastic tensor can be supplied by the user. With the compliance $S=C^{-1}$, the difference between the internal stress and a user-supplied target stress $\sigma_{\mathrm{ext}}$ can be used to calculate a strain,
\begin{equation}
\varepsilon_{\rho} = -\,S_{\rho\xi}\,\bigl(\sigma_{\mathrm{int},\xi}-\sigma_{\mathrm{ext},\xi}\bigr),
\label{eq:strain}
\end{equation}
an engineering Voigt strain whose shear components are halved when it is written back as a symmetric $3\times3$ tensor.
Before it is applied, the strain is projected onto the subspace allowed by the symmetry of the current cell. This symmetry is the point group $\mathcal{P}$ of the space group, which contains the non-translational part of every operation of the space group. With $Q$ denoting these operations in Cartesian coordinates, the projection is the group average
\begin{equation}
\bar{\varepsilon} = \frac{1}{|\mathcal{P}|}\sum_{Q\in\mathcal{P}} Q\,\varepsilon\,Q^{\mathsf T}.
\label{eq:strainproj}
\end{equation}
By Neumann's principle \cite{nye1985}, $\bar{\varepsilon}$ contains only the strain components a crystal of that symmetry may have: one for a cubic cell, two for hexagonal, trigonal and tetragonal, three for orthorhombic, four for monoclinic and six for triclinic. The removed part, $\lVert\varepsilon-\bar{\varepsilon}\rVert$, is the symmetry-forbidden component of the stress residual, which is pure sampling noise (the cell counterpart of the Wyckoff residual $\varrho_a$); it is monitored at every lattice step.

The projected strain is applied to the lattice matrix,
\begin{equation}
\mathscr{A}^{n+1} = \mathscr{A}^{n}\,(\mathcal{I} + \lambda_l\varepsilon),
\end{equation}
where $\lambda_l$ is a damping factor. Analogous to $\lambda_{wd}$, $\lambda_{l}$ is a base value $\lambda_0$, supplied by the user or unity, times a noise-referenced gain, $\lambda_l = \lambda_0\,z^2/(z^2+z_0^2)$, where $z$ is the largest component of the stress residual in units of its standard error over the sampled ensemble, and $z_0=3$. Far from convergence, the full step is taken; once the residual is within a few standard errors of zero, the step shrinks smoothly, so that the last steps average over several sample sets instead of transferring the canonical sampling noise of the stress to the cell. 

The user provides a target pressure. The target is given as pressures, positive in compression, and converted to stresses in the tension-positive convention. This way $\sigma_{\mathrm{ext}}$ is supplied by the user and may take three forms: hydrostatic, with equal diagonal components; non-hydrostatic diagonal, with different normal stresses along the three axes; or a general non-diagonal tensor, including shear, i.e. $\sigma_{\mathrm{ext},i}=-P_i$; a hydrostatic pressure $P$ thus corresponds to $\sigma_{\mathrm{ext}}=-P\,\mathbb{I}$.

\subsection{The relaxation workflow}
\label{sec:workflow}

The loop shown in Fig.~\ref{fig:flowchart} consists of four stages.

\paragraph{Start.} A static variable-cell relaxation at the target pressure supplies the starting structure.

\paragraph{Preconditioning.} Initial model force constants are generated from an estimate of the maximum phonon frequency of the system or from its Debye temperature \cite{TDEP_s-tdep}. A small ensemble is sampled from these force constants, the forces are computed, and the force constants are refitted. The ensemble is then enlarged on a user-defined schedule, typically a geometric series such as 4, 8, 16, 32, 64 configurations chosen per system. In this schedule the samples of the previous iteration are also used in the fit of the current iteration. This is repeated  until the phonon spectrum has no imaginary frequencies and the second-order fit reaches a cross-validation score $R^2\geq0.6$. If neither is reached by the end of the schedule, the run stops. From the third ensemble of the schedule on, the positions are updated, provided the lowest phonon frequency is above $-1$~THz.

\paragraph{Positions.} A user-defined number of configurations is sampled from the current force constants and their forces are computed. The force constants are fitted order by order: the second-order force constants first, then the first-order force constants $\Phi^{(1)}$ from the residuals, Eq.~\eqref{eq:fc1}, and finally the third-order force constants from the residual that remains after both second and first order contributions are subtracted. The Wyckoff-projected displacement is then applied. These steps repeat until the mean and maximum of $|\Phi^{(1)}|$ are below a chosen tolerance. If the root-mean-square change of the fractional coordinates in a step is below a threshold, the samples of that step are merged into the next fit, so that the ensemble accumulates while the structure changes little; otherwise the next step is fitted on fresh samples only. The first position step after a lattice update always uses fresh samples.

\paragraph{Lattice.} The force constants are refitted on the accumulated ensemble, the phonon contribution to the stress and the compliance tensor are computed, and the cell is updated by the damped, symmetry-projected strain. The lattice loop has converged when three conditions hold together: (i)~$\max_i|\sigma_{\mathrm{int},i}-\sigma_{\mathrm{ext},i}|$ is below a chosen pressure tolerance; (ii)~the change in the phonon contribution to the stress between successive lattice steps is below the same tolerance; and (iii)~the relative volume change of the cell in the preceding lattice step, $|\Delta V/V|$, is below 1\%. Conditions (ii) and (iii) guard against a step that merely happens to cross the target. Optionally, to avoid demanding a convergence criterion below the sampling noise, the option is there to raise the tolerance for each stress component is  to $K$ times its standard error over the sampled ensemble, given that this new condition is larger than the originally requested tolerance. 

Checkpoints are written after every stage, so a run interrupted at \textbf{any} step resumes from the last completed stage.

\begin{figure*}[t]
    \centering
    \includegraphics[width=0.9\linewidth]{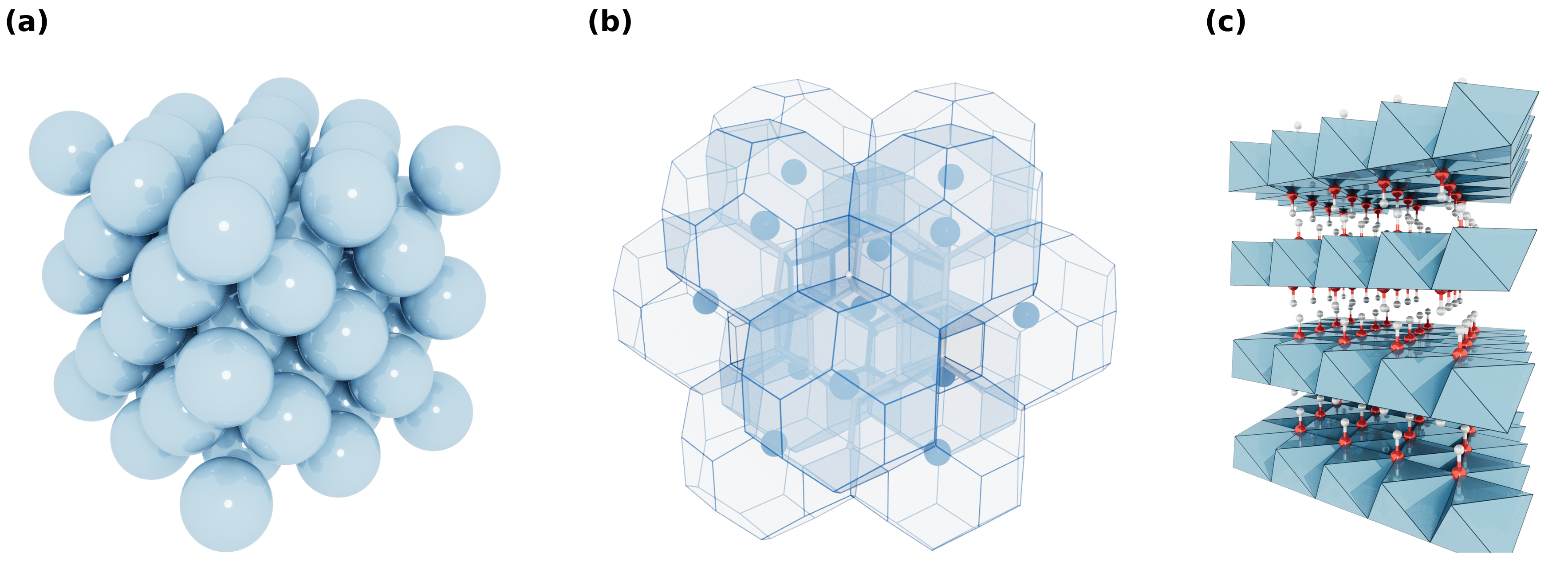}
    \caption{\justifying \textbf{The three test systems as relaxed with \warp{}.} (a) Aluminium, fcc, a block of $2\times2\times2$ conventional cells. (b) LaH$_{10}$ in its cubic $Fm\bar{3}m$ phase at 150~GPa and 300~K, with the La-centred H$_{32}$ cages. (c) Brucite, Mg(OD)$_2$, at 300~K: four layers of edge-sharing MgO$_6$ octahedra, with the O--D groups pointing into the gap between them. Rendered with CrysViz (\url{https://crysviz.org/}).}
    \label{fig:systems}
\end{figure*}
\subsection{Machine-learned interatomic potentials}
\label{sec:mlip}
Even though sTDEP is highly efficient, running several tens or hundreds of supercell calculations per (\textit{T,P}) point can become very computationally expensive. Especially when the physical effects over a range of temperatures and pressures are evalauted. Since \warp{}  requires only the forces and stresses, one can replace the DFT engine with an MLIP. In \warp{} the force engine is interfaced through the Atomic Simulation Environment \cite{larsen2017ase}, so any calculator can be used. The potentials used in this work are UPET models \cite{bigi2026upet}, foundation models of the PET architecture \cite{mazitov2025petmad} pre-trained on the OMat24 dataset \cite{barrosoluque2024omat24} and fine-tuned on DFT reference configurations of the system at hand. The training sets, splits, fine-tuning settings and held-out errors of every potential are given in Supplementary Section~S4.

\section{Results}
As \warp{} provides us with new and efficient routes to structures at finite temperature and high pressure conditions, we wanted to test the capabilities of this method. The three systems (Fig.~\ref{fig:systems}) test different parts of the method. In face-centred cubic aluminium, every atom sits on a site without a free Wyckoff parameter, so the position update returns no displacement and only the cell is relaxed. The hydrogen sites of the lanthanum hydrides do have free parameters, and in the LaH$_{10}$ transition of Sec.~\ref{sec:lah_mlip} these parameters converge to values at which the rhombohedral hydrogen sites become the $8c$ and $32f$ sites of $Fm\bar{3}m$; this is the most direct test of the position projection. In brucite, oxygen and deuterium each sit on the three-fold axis with one free parameter, and the O--D distance of Sec.~\ref{sec:brucite} depends on these two parameters and on the $c$ lattice parameter, so it tests the position and cell updates together. In all three cases, convergence is judged on the residual forces and the stress at the resolution of the sampled ensemble.

\begin{figure}[t]
    \centering
    \includegraphics[width=\linewidth]{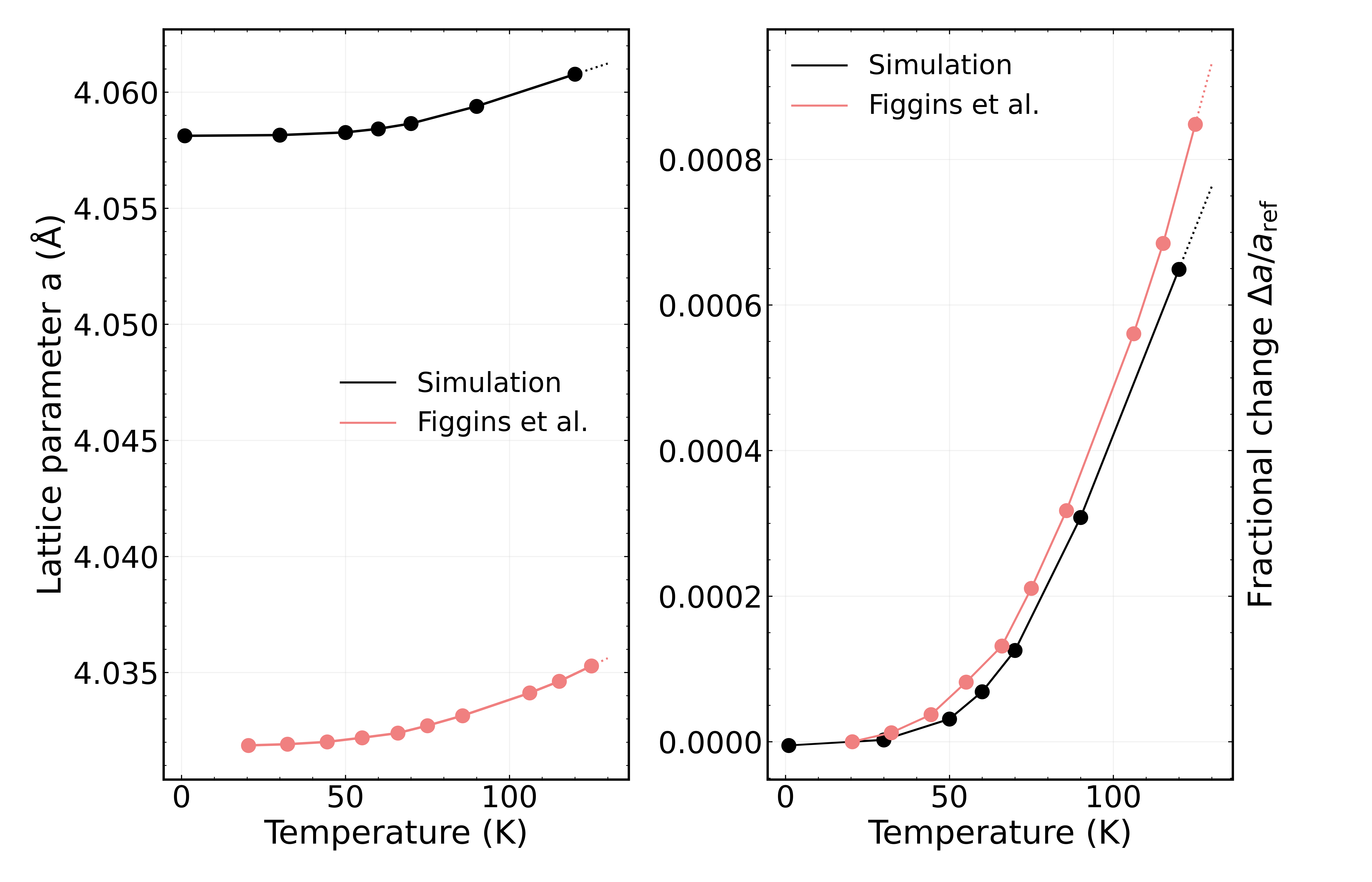}
    \caption{\justifying \textbf{Aluminium at low temperature.} Left: absolute lattice parameter. Right: fractional change relative to the lowest experimental point at 20.4~K \cite{Al_down_to_20K_thermal_expansion};  Both series end near 120~K, and the dotted continuations at the right-hand edge of each panel carry no data. The curvature below 100~K is reproduced.}
    \label{fig:al_lowT}
\end{figure}
\begin{figure*}[t]
    \centering
    \includegraphics[width=0.8\linewidth]{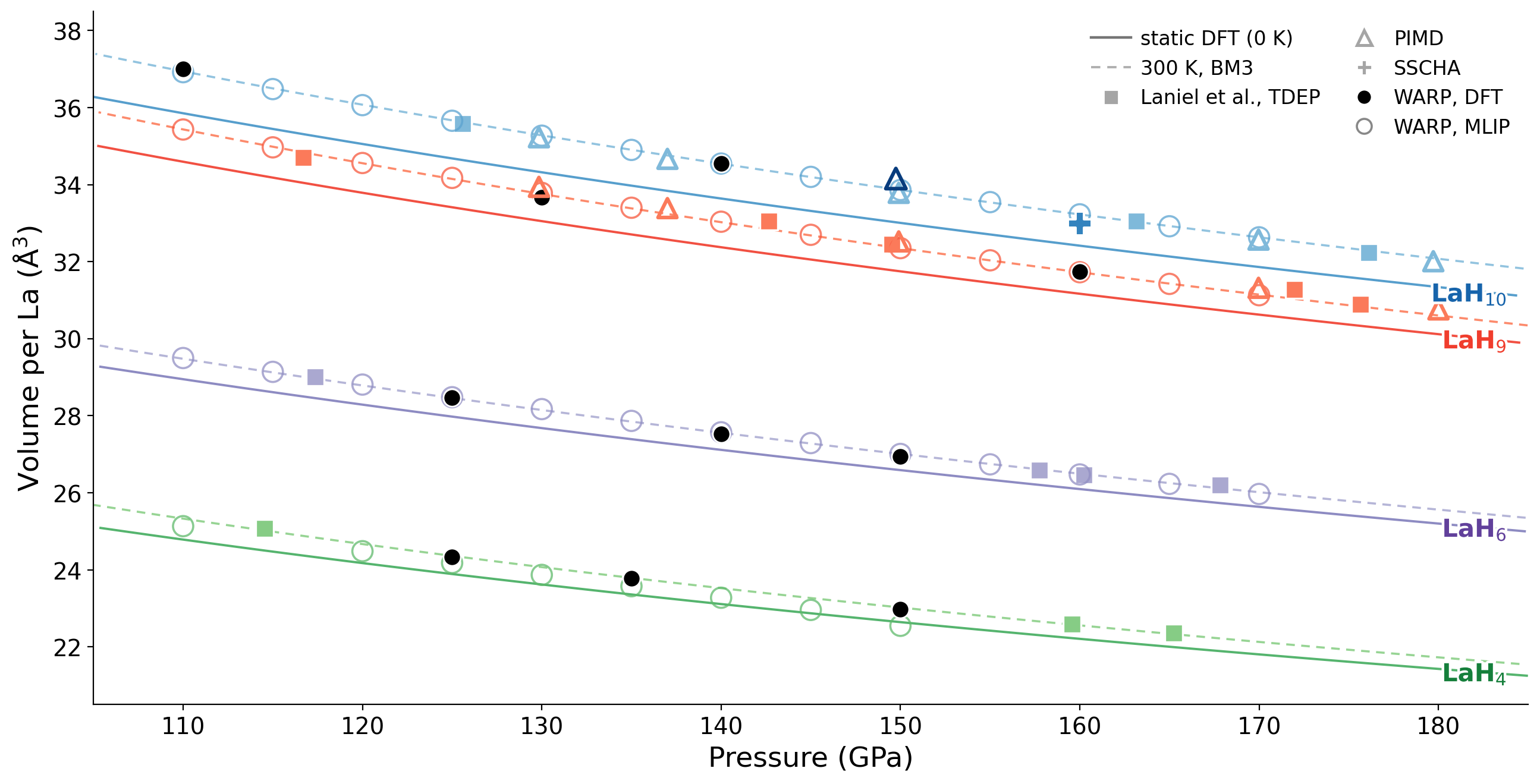}
    \caption{\justifying \textbf{Pressure--volume relations of lanthanum hydrides at 300~K.} Squares: TDEP calculations of  \citet{laniel2022high}. Open triangles: PIMD of \citet{wang2023quantum,PIMD_Ackland}. Crosses: SSCHA of \citet{errea2020quantum}. Filled black circles: \warp{} with DFT forces. Open circles: \warp{} with the transfer-learned potential; LaH$_4$ and the pressures below 118~GPa were not in its training set. There are ten filled circles, three for LaH$_4$, three for LaH$_6$ and two each for LaH$_9$ and LaH$_{10}$; where both were run at the same pressure the filled circle is drawn over the open one. PIMD triangles are drawn in front. Solid lines: the static (0~K) DFT Birch--Murnaghan equations of state. Dashed lines: Birch--Murnaghan curves through the 300~K points, the \warp{} DFT-force points together with the TDEP points of \citet{laniel2022high} for LaH$_4$ and LaH$_6$, and the \warp{} DFT-force and potential points for LaH$_9$ and LaH$_{10}$, with $K_0'$ held at the static value; the legend shortens this to a fit through the \warp{} points. They are guides for reading the PIMD and SSCHA points against the finite-temperature result rather than the static one.}
    \label{fig:LaHx_VP}
\end{figure*}
\subsection{Thermal expansion of aluminium}

Aluminium is one of the most thoroughly characterised metals, with reliable thermal expansion data to cryogenic temperatures \cite{Al_down_to_20K_thermal_expansion}, and its use in aerospace structures makes its low-temperature dimensional stability of direct engineering interest \cite{tunes2025legacy}. Figure~\ref{fig:al_lowT} shows the comparison of the calculated lattice parameter with the measurements of \citet{Al_down_to_20K_thermal_expansion} from 20.4~K to around 120~K at $P=0$, where the measured series ends. The fractional change $\Delta a/a_{20.4\,\mathrm{K}}$ follows the measurement, including the curvature below 100~K that a classical treatment does not contain \cite{TDEP_Si}. Towards higher temperatures the calculated fractional difference falls below the measured expansion. The absolute lattice parameter sits 0.026~\AA, or 0.65\%, above the measurement, and that offset is close to constant across the range. 

\begin{figure}[t]
\centering
\resizebox{\linewidth}{!}{%
\begin{tikzpicture}[
  font=\small, >=Latex,
  box/.style={draw, rounded corners=2pt, line width=0.9pt, align=center, minimum height=11mm, minimum width=27mm, inner sep=3pt},
  grey/.style={box, draw=black!45},
  dft/.style={box, draw=blue!60!black},
  ml/.style={box, draw=green!50!black},
  arr/.style={->, line width=0.9pt, black!55},
  ft/.style={->, line width=1.1pt, violet!70!black},
  lbl/.style={font=\footnotesize, align=left, text=black!70},
]
\def\dx{50mm}\def\dy{15mm}
\node[grey] (s1) at (0,0) {\textbf{Start}\\[1pt]{\scriptsize static 0~K minimum}\\{\scriptsize at target $P$, DFT}};
\node[dft]  (p1) at (0,-\dy) {\textbf{Precondition}\\[1pt]{\scriptsize sample $\to$ fit}\\{\scriptsize until phonons real}};
\node[dft]  (x1) at (0,-2*\dy) {\textbf{Positions}\\[1pt]{\scriptsize $\Phi^{(1)}\to$ Wyckoff-}\\{\scriptsize projected update}};
\node[dft]  (l1) at (0,-3*\dy) {\textbf{Lattice}\\[1pt]{\scriptsize $\varepsilon=-C^{-1}\Delta\sigma$}};
\node[grey] (f1) at (0,-4*\dy) {\textbf{Converged}\\[1pt]{\scriptsize structure at $(P,T)$}};
\foreach \u/\v in {s1/p1,p1/x1,x1/l1,l1/f1} \draw[arr] (\u)--(\v);
\draw[arr] (l1.west) .. controls +(-9mm,0) and +(-9mm,0) .. (x1.west);
\node[lbl, anchor=south west] at (-14mm,8mm) {\textbf{a}\quad DFT forces};
\node[ml]   (ftn) at (\dx,0) {\textbf{Fine-tune}\\[1pt]{\scriptsize foundation MLIP on}\\{\scriptsize the supercells from (a)}};
\node[grey] (s2)  at (\dx,-\dy) {\textbf{Start}\\[1pt]{\scriptsize static 0~K minimum}\\{\scriptsize \emph{of the fine-tuned MLIP}}};
\node[ml]   (p2)  at (\dx,-2*\dy) {\textbf{Precondition}\\[1pt]{\scriptsize MLIP phonons real}};
\node[ml]   (x2)  at (\dx,-3*\dy) {\textbf{Positions}};
\node[ml]   (l2)  at (\dx,-4*\dy) {\textbf{Lattice}};
\node[grey] (f2)  at (\dx,-5*\dy) {\textbf{Converged}\\[1pt]{\scriptsize one core, hours}};
\foreach \u/\v in {ftn/s2,s2/p2,p2/x2,x2/l2,l2/f2} \draw[arr] (\u)--(\v);
\draw[arr] (l2.east) .. controls +(9mm,0) and +(9mm,0) .. (x2.east);
\draw[ft] (p1.east) -- node[midway, fill=white, inner sep=1.5pt, font=\footnotesize, text=violet!70!black]{training set} (ftn.west);
\node[lbl, anchor=south west] at (\dx-14mm,8mm) {\textbf{b}\quad MLIP forces};
\end{tikzpicture}}%
\caption{\justifying \textbf{The two ways to run the relaxation.} (a) With DFT forces, every sampled supercell is a first-principles calculation. The loop consists of preconditioning, Wyckoff-projected position updates, and lattice updates through the symmetry-projected strain, until the internal stress matches the target (details in Fig.~\ref{fig:flowchart}). (b) The supercells sampled in (a) form the reference set for fine-tuning a foundation potential. The relaxation then restarts from the beginning: since every quantity must be consistent with the force engine used throughout, the reference structure is first re-relaxed to the static 0~K minimum of the fine-tuned potential itself, after which the same loop runs at any temperature and pressure. For LaH$_{10}$ at 150~GPa, the 150 DFT supercells taken from the first three iterations of the DFT-force run were the entire first-principles cost.}
\label{fig:pipeline}
\end{figure}
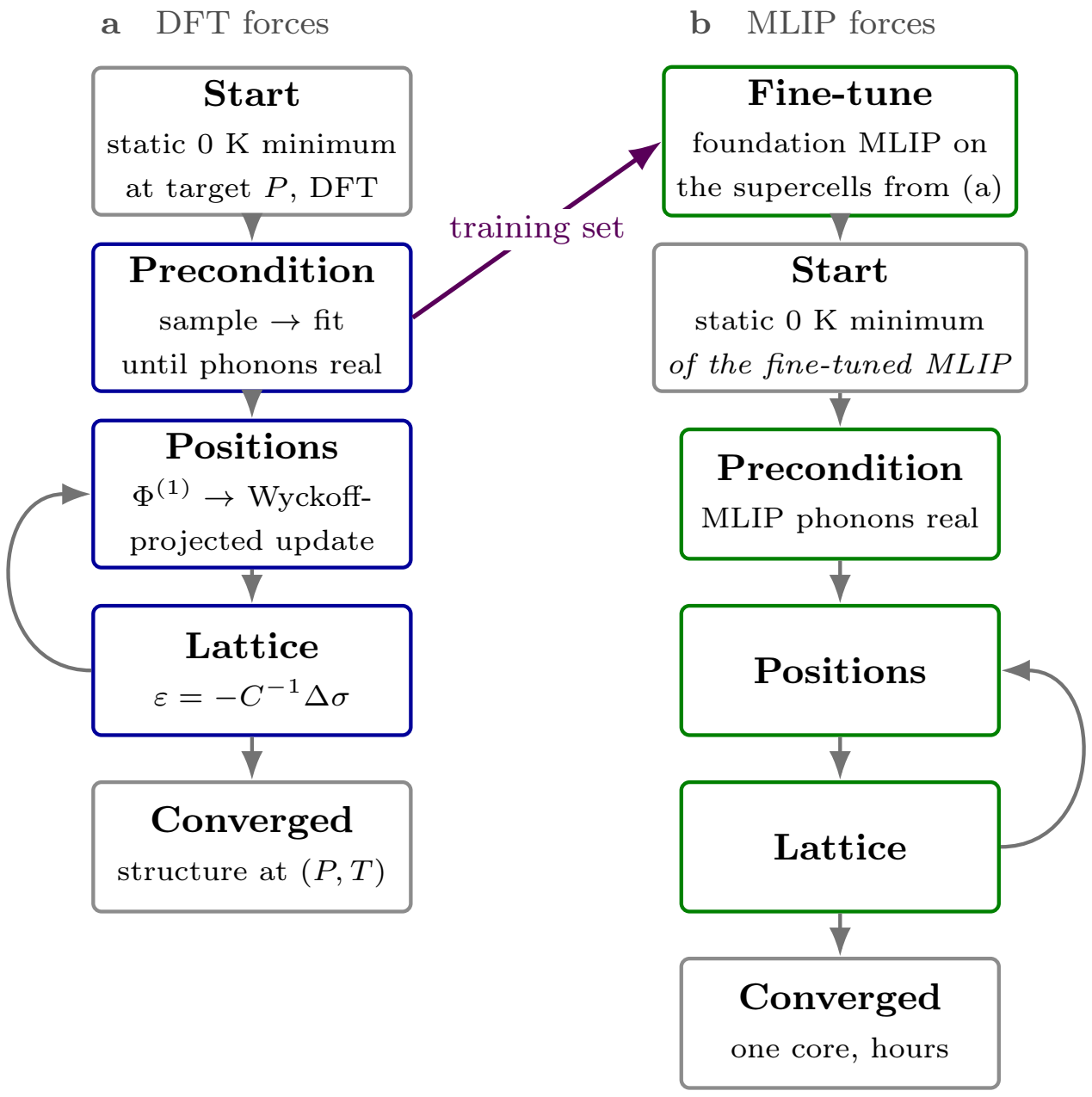
\subsection{Lanthanum hydrides at megabar pressure: equations of state}
\label{sec:lah_dft}

\begin{figure*}[t]
    \centering
    \includegraphics[width=0.85\linewidth]{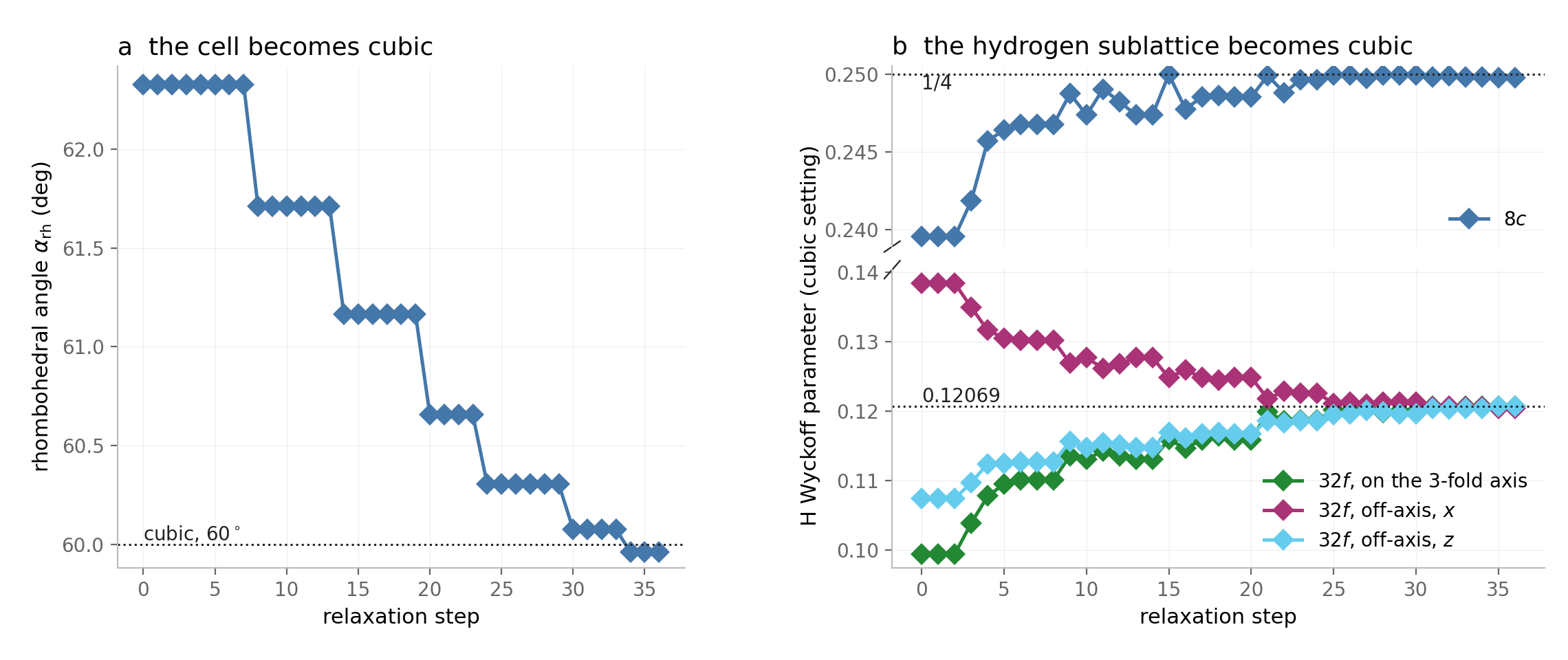}
\caption{\justifying \textbf{LaH$_{10}$ at 150~GPa and 300~K: $R\bar{3}m\to Fm\bar{3}m$ with DFT forces.} (a) Rhombohedral angle of the primitive cell against relaxation step, counting every preconditioning, position and lattice step. Seven plateaux, and therefore six lattice steps, take it from $62.327^\circ$ to $59.962^\circ$. (b) Hydrogen Wyckoff parameters in the cubic setting. The $8c$ parameter, measured from the La site, reaches $\tfrac14$. The three parameters of the $32f$ orbit, one for the pair on the three-fold axis and two for the six off-axis atoms, measured from the cage centre, merge into the single value 0.1207 of the one free parameter of the cubic $32f$ site. Compare Extended Data Fig.~8 of \citet{errea2020quantum}.}
    \label{fig:lah_transition}
\end{figure*}
Lanthanum hydrides are hydrogen-rich, strongly anharmonic, at megabar pressure, and known to be poorly described by static DFT, with finite-temperature and nuclear-quantum effects narrowing but not closing the gap to experiment \cite{laniel2022high}. Figure~\ref{fig:LaHx_VP} compares the \warp{} calculated pressure--volume relations for LaH$_4$, LaH$_6$, LaH$_9$ and LaH$_{10}$ at 300~K with the TDEP results of \citet{laniel2022high}, PIMD by \citet{wang2023quantum} and by \citet{PIMD_Ackland}, and SSCHA by \citet{errea2020quantum}. The volumes as calculated with \warp{} are in excellent agreement across all four stoichiometries and the full pressure range, over volume changes of more than ten per cent, with 16--40 configurations per position step in cells of up to 231 atoms.

\label{sec:lah_mlip}
As previously described, the large number of DFT supercell calculations needed for a full P-V equation-of-state (EOS) at a given temperature can be computationally demanding if large (\textit{T,P}) ranges are explored and analyzed. For this reason we evaluated the use of MLIPs. Every supercell sampled in the DFT-force relaxations is a first-principles calculation, and these supercells can serve as the training set for a potential used in the same relaxation (Fig.~\ref{fig:pipeline}). Several benchmarks were performed in order to investigate the efficient of training a potential from the DFT samples. Initially all 3383 samples from the full DFT-based simulations were used to fine-tune a UPET model \cite{bigi2026upet,mazitov2025petmad}, containing LaH6, LaH9, and LaH10 structures over a wide pressure range, as described in the methods section. Two extrapolations were built into the test deliberately; i) LaH$_4$ is absent from the training set, and ii) the pressure range extends below the lowest trained pressure. The DFT based calculations supply ten reference volumes across the four stoichiometries, and at every tested pressure, the potential reproduces the reference volume to within 0.4\%: with the largest deviation at 150~GPa, ranging from 0.27\% for LaH$_6$, to 0.35\% for LaH$_9$. For LaH$_{10}$, the agreement is 0.01\% at 140~GPa (34.547 against 34.549~\AA$^3$ per La) and 0.2\% at 110~GPa (36.919 against 36.998~\AA$^3$ per La).  The largest error is 1.87\% for the out-of-distribution LaH$_4$,

\begin{figure}[t]
    \centering
    \includegraphics[width=\linewidth]{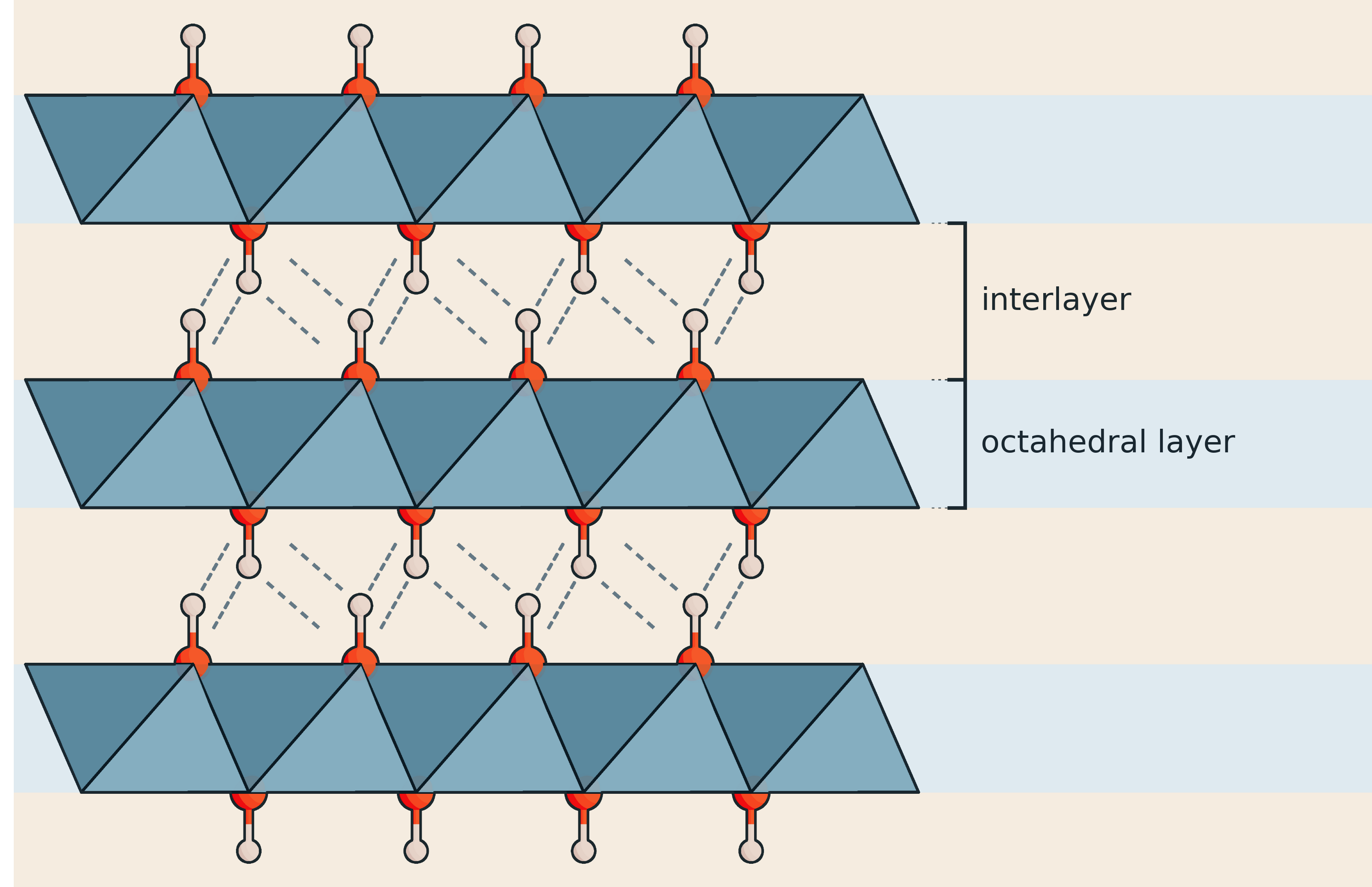}
    \caption{\justifying \textbf{Brucite Mg(OD)$_2$ and the two parts into which its thermal expansion is resolved.} The structure at 300~K relaxed with the PBE$+$D3 potential, viewed along $a$ with $c$ vertical: edge-sharing MgO$_6$ octahedra (blue), oxygen (red) and deuterium (white), with the shortest D$\cdots$O contacts across the gap dashed. The octahedral layer is measured between the two oxygen planes of one sheet and the interlayer between the oxygen planes on either side of the gap; the shaded bands mark the two regions in every layer.}
    \label{fig:brucite_layers}
\end{figure}
\begin{figure*}[t]
    \centering
    \includegraphics[width=0.8\linewidth]{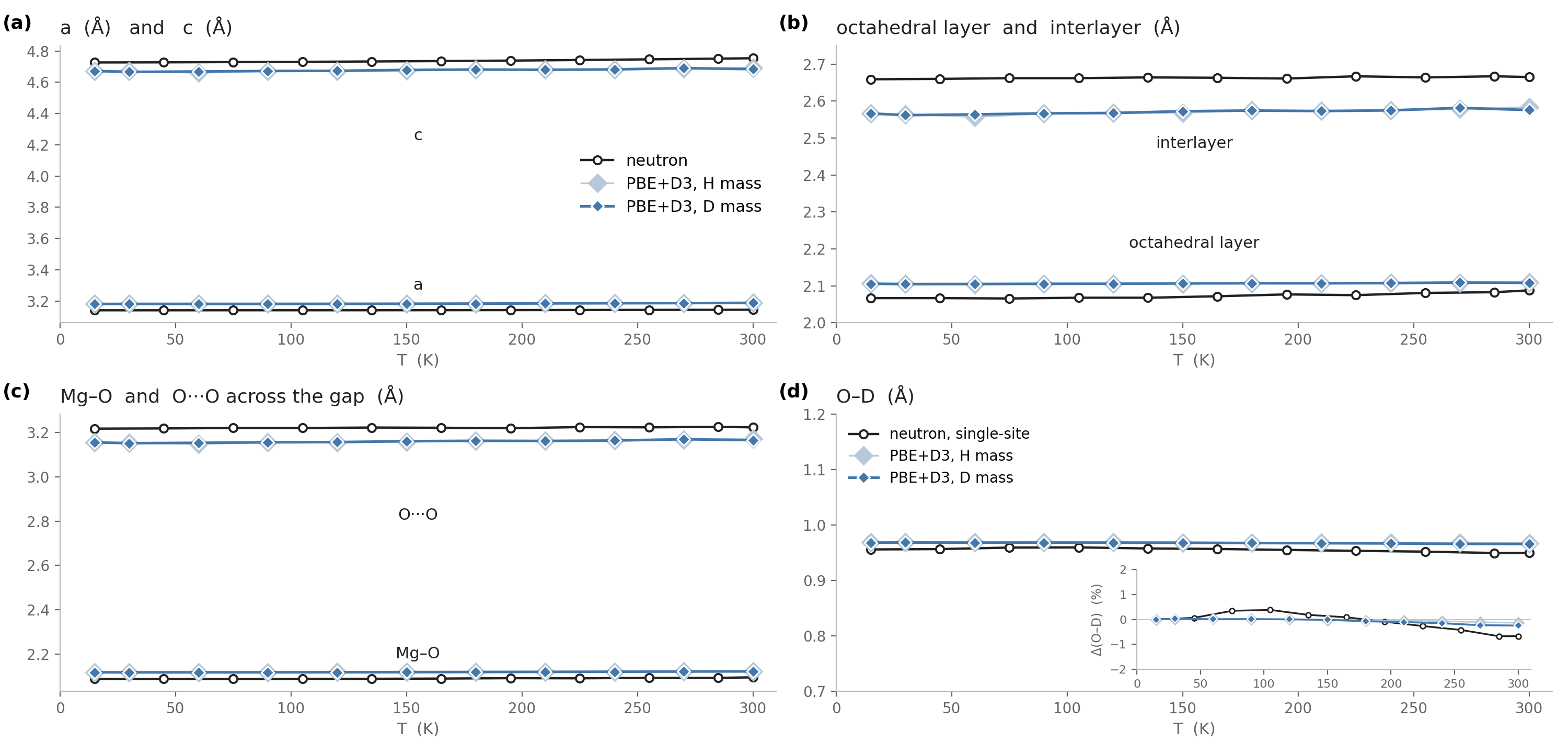}
    \caption{\justifying \textbf{Brucite Mg(OD)$_2$ at ambient pressure against neutron diffraction.} This work with the PBE$+$D3 potential and the deuterium mass on the hydrogen sites (dark blue diamonds, 15~K and 30--300~K in 30~K steps), with the protium-mass series behind it (light blue diamonds), against the neutron diffraction data of \citet{chakoumakos1997brucite} (open black symbols). Every computed point is the final structure of its relaxation, the last lattice step. The lines join the computed points. (a) The lattice parameters $a$ (lower group) and $c$ (upper group) on one axis. (b) The octahedral-layer thickness (lower group) and the interlayer thickness (upper group). (c) The Mg--O bond (lower group) and the O$\cdots$O distance across the interlayer gap (upper group). (d) The O--D distance against the single-site model of the neutron data (open circles); the inset shows the change of each series relative to its value at the lowest temperature, in percent. The computed O--D is the distance between the mean oxygen and deuterium positions with the deuterium on the three-fold axis, which is the quantity the single-site model gives.}
    \label{fig:brucite}
\end{figure*}
LaH$_{10}$ at 150 GPa was further investigated in order to benchmark whether the method captures finite-temperature phase transitions correctly. Static PBE relaxation result in a rhombohedral LaH$_{10}$ structure, whereas the experimentally observed structure is cubic ($Fm\bar{3}m$). \citet{errea2020quantum} showed with SSCHA that quantum nuclear motion stabilises the cubic phase \cite{errea2020quantum}. In their approach they used up to 1000 configurations per population in an 88-atom supercell. Starting from the rhombohedral structure, \warp{} reproduces this quantum stabilisation with 40 configurations per position step in a 231-atom cell . In six lattice steps, the rhombohedral angle reduces from $62.327^\circ$ to $59.962^\circ$, compared to $60^\circ$ for the cubic cell, and the hydrogen $8c$ parameter reaches $\tfrac14$, the value of the special Wyckoff position in the cubic space group. The $32f$ site keeps one free parameter in $Fm\bar{3}m$, so its signature of the transition is not a fixed value but the merging of its three rhombohedral parameters into a single one, 0.1207 (Fig.~\ref{fig:lah_transition}). No symmetry was imposed on the result, only on each step. A \warp{} relaxation confined to $R\bar{3}m$ can therefore converge efficietly into $Fm\bar{3}m$ without the cubic symmetry being imposed.

To test how little DFT data is needed, the foundation model was fine-tuned on only 150 supercells taken from the first three lattice steps of the DFT-force run of LaH$_{10}$, 120 structures were used for training, 15 for validation and 15 for testing. None of these supercells were close to the cubic endpoint, yet the relaxation with this potential reproduces the $R\bar{3}m\to Fm\bar{3}m$ transition (Supplementary Fig.~S1). The 150 DFT supercells are therefore the entire first-principles cost of the complete \warp{} relaxation (Fig.~\ref{fig:pipeline}b).

\subsection{Brucite Mg(OD)$_2$: Direct comparison with neutron diffraction data}
\label{sec:brucite}

The hydrogen positions of the lanthanum hydrides could only be compared with other calculations. In brucite, however, the hydrogen positions have been measured directly. It is stable at ambient pressure and the structure of deuterated samples has been measured by neutron diffraction from 15 to 300~K \cite{chakoumakos1997brucite}, making it a model for the hydrous minerals of planetary interiors \cite{hermann2016brucite,schaack2020brucite}. Mg(OD)$_2$ is a layered hydroxide in which sheets of octahedral MgO are separated by a gap. Every O--D unit is oriented towards this gap. Its structure therefore contains two kinds of distances between atoms, those within a sheet of polyhedra and those across the gap (Fig.~\ref{fig:brucite_layers}). The crystal structure is in space group $P\bar{3}m1$, where O and D each occupy the three-fold axis with one free parameter $z$, the computed O--D distance, $(z_{\mathrm D}-z_{\mathrm O})\,c$, is a distance between mean positions. \citet{chakoumakos1997brucite} modelled the deuterium atoms both on a single site on the axis and over three sites displaced from it; the uncorrected O--D distance of the single-site model is given by the same expression and is the quantity compared here.

Mg(OD)$_2$ was relaxed at ambient pressure from 15 to 300~K, with the UPET potential \cite{bigi2026upet} fine-tuned on thermally displaced supercells from PBE$+$D3 DFT based calculations (Supplementary Section~S4); it reproduces the static PBE$+$D3 cell to 0.3\% or better (Supplementary Section~S2). The values at every temperature are taken from the last lattice step of each relaxation and are listed in Supplementary Table~S2. Because the thermal expansion of brucite between 15 and 300~K is only 0.027~\AA\ in $c$, every lattice step averages the stress over 1024 configurations of the 215 atom cells. The 512 samples of the position step are included in the lattice step. The standard error of $\sigma_{zz}$ is 0.07--0.09~GPa, which, with using the measured elastic constants \cite{jiang2006brucite}, corresponds to around 0.008~\AA\ (0.2\%) in $c$ and around 0.0002~\AA\ in $a$; the octahedral layer and the O--D distance, which follow from the positions, are resolved to around 0.001 and 0.0003~\AA.

The cell and the bond distances agree with experimental values over the complete temperature range (Fig.~\ref{fig:brucite}a--c): $a$ is 1.3 to 1.4\% longer and $c$ 1.2 to 1.5\% shorter than measured, the Mg--O bond 1.3 to 1.4\% longer and the octahedral layer 1.0 to 1.9\% thicker. The gap is slightly compressed, the interlayer distance being 3.2 to 3.7\% and the O$\cdots$O distance across it 1.7 to 2.1\% shorter, an offset already present in the static cell (Supplementary Section~S2).

The thermal expansion of the cell is in good agreement with experimental data: between 15 and 300~K the volume increases by 0.70\%, against 0.79\% measured, and $c$ by 0.012~\AA, compared to 0.027~\AA\ in measurements. Its partition between the octahedral layer and the interlayer distance, however, is reversed. In the calculation the increase of $c$ is carried by the interlayer distance (0.009~\AA, against 0.003~\AA\ for the octahedral layer), whereas in experiment it is carried by the octahedral layer (0.021~\AA, against 0.006~\AA\ for the interlayer distance), although the measured interlayer distance is not monotonic and its total change is only around twice its scatter between adjacent temperatures. The series with the protium mass shows the same partition (Supplementary Section~S2). Since the relaxation treats the sheets and the gap on the same footing, the likely origin is the description of the forces across the gap, which are weak and dominated by dispersion: the thermal response of the interlayer distance is most likely not well captured by current density functionals. Establishing this requires a dedicated study of the interlayer distance, beyond the scope of the presentation of this method.

The computed O--D distance is 0.009 to 0.016~\AA, or 0.9 to 1.7\%, longer than the measured one (Fig.~\ref{fig:brucite}d), and its temperature dependence shows a clear isotope effect. With the protium mass on the hydrogen sites the distance decreases by 0.0015~\AA\ between 30 and 300~K. With the deuterium mass, which leaves the potential unchanged and alters only the amplitudes of the sampled motion, the decrease nearly doubles, to 0.0026~\AA, against 0.0067~\AA\ measured over the same range. The origin is the libration of the O--D unit, the motion of the deuterium perpendicular to the bond, which shortens the distance between the mean positions of oxygen and deuterium. The libration of the heavier deuteron lies lower in frequency and is thermally excited at lower temperature, so its amplitude, and with it the shortening, grows faster with temperature than for the proton, although at 300~K both are still dominated by zero-point motion. The measured distance first rises to a maximum of 0.9599~\AA\ at 105~K and then falls; neither series shows this maximum.

\section{Discussion}

The central design choice of \warp{}, taking every step in the Wyckoff parameters and symmetric strains of the current space group, proves to be more than a convenience. It removes the symmetry-forbidden part of the sampling noise by construction, and, because the structures of a supergroup form a subspace of the parameter space of the subgroup, it allows the symmetry to rise when the free energy favours it. LaH$_{10}$ shows both: starting from the rhombohedral static minimum, the relaxation reaches the cubic phase without the transition being imposed, reproducing the quantum stabilization found with SSCHA by \citet{errea2020quantum}, with 40 configurations per position step. Weighing the symmetry-allowed stress against its own standard error then lets the relaxation stop at the resolution of the ensemble rather than at an arbitrary tolerance. The resulting parameters agree perfectly with the results by \citet{errea2020quantum}, see Extended Data Fig.~8 of \citet{errea2020quantum}.

Where the results differ from experiment, they do so in ways that can be traced to the force engine rather than to the relaxation. The absolute lattice parameter of aluminium is 0.026~\AA\ too large, while its fractional thermal expansion follows the measurement at cryogenic temperatures. In brucite, the cell, the bond distances and the thermal expansion agree with experiment, while the partition of the expansion between the octahedral layer and the interlayer distance points to the description of the interlayer by current density functionals. This makes \warp{} a tool for assessing force engines at finite temperature: with anharmonicity and zero-point motion included, the remaining differences from experiments can be traced to the functional or the potential.

The method has clear limits, each of which suggests its own extension. Because the projection lets symmetry rise but never fall, a structure that has reached the subspace of a supergroup stays there even if this is a saddle point of the free energy. The ensemble is a single Gaussian centred on the mean positions, which captures libration but not a distribution over several sites, such as the off-axis disorder of the deuterium in brucite; this part of the temperature dependence of the O--D distance remains outside the method. Finally, the lattice step is preconditioned by the fitted compliance tensor, which describes soft, long-ranged responses along the interlayer direction of brucite less well, so that the cell converges in tens of steps rather than a few. For such cases, \warp{} accepts a full elastic tensor supplied by the user, for instance from experiment, in place of the fitted one; since the compliance tensor only sets the step, this changes the path of the relaxation but not its fixed point.

\section{Conclusion}

\warp{} allows the relaxation of crystal structures at finite temperature and pressure on the effective Hamiltonian of sTDEP, with anharmonicity and zero-point motion included and every step confined to the symmetry of the current space group. It follows the measured thermal expansion of aluminium including its low-temperature quantum plateau, reproduces the equations of state of four lanthanum hydrides at megabar pressures, and finds the cubic phase of LaH$_{10}$ from its rhombohedral static minimum, with DFT forces and with a potential fine-tuned on 150 of its own supercells. In brucite, a layered hydrous mineral, it places the deuterium atoms to within 0.016~\AA\ of the measured O--D distance, picks up the isotope effect on that distance, and reproduces the cell and the Mg--O bond to within 1.5\% from 15 to 300~K. Using density-functional or machine-learned forces through a common interface, \warp{} makes finite-temperature structures accessible for systems ranging from metals and super-hydrides to hydrous minerals relevant to planetary interiors.%

\section*{Data availability}
The relaxed structures, force constants, training sets and fine-tuned potentials underlying the figures are available from the corresponding author on request.

\section*{Code availability}
\warp{} is available in the repository \url{https://github.com/TEPEC-Physics-at-Extreme-Conditions/WARP}.

\section*{Acknowledgements}
We thank Rickard Armiento (Link\"oping University) for the tabulation of the space-group Wyckoff degrees of freedom on which the projector is built (\url{https://symdata.anyterial.se/}), and Gerd Steinle-Neumann (Bayerisches Geoinstitut, University of Bayreuth) for suggesting brucite as a test case. F.K. acknowledges support from the Swedish Research Council (VR) grant 2020-04630 and the Swedish e-Science Research Centre (SeRC). H.R.A.t.E.\ and F.T.\ acknowledge support through ERC Grant (UNMASCC-HP, 10.3030/101117758). Funded by the European Union. Views and opinions expressed are however those of the author(s) only and do not necessarily reflect those of the European Union or the European Research Council Executive Agency. Neither the European Union nor the granting authority can be held responsible for them. F.T.\ acknowledges furthermore support from the Swedish government's Strategic Research Area in Materials Science on Functional Materials at Link\"oping University (Faculty Grant SFO-Mat-LiU 2009-00971). Computations were enabled by resources provided by the National Academic Infrastructure for Supercomputing in Sweden (NAISS) on Dardel at PDC, partially funded by the Swedish Research Council through grant agreement no.~2022-06725. We acknowledge NAISS for providing access to the LUMI supercomputer, owned by the EuroHPC Joint Undertaking and hosted by CSC (Finland) and the LUMI consortium. For open access, the authors have applied a Creative Commons Attribution (CC BY) license to any Author Accepted Manuscript version arising from this submission.

\section*{Author contributions}
\textbf{H.R.A.t.E.}: Conceptualization, Methodology, Software, Investigation, Formal analysis, Visualization, Writing -- original draft. \textbf{F.T.}: Conceptualization, Funding, Supervision, Writing -- review \& editing. \textbf{F.K.}: Conceptualization, Writing -- review \& editing.

\section*{Competing interests}
The authors declare no competing interests.

\onecolumngrid
\clearpage
\hypersetup{pageanchor=false}
\setcounter{footnote}{0}
\makeatletter
\let\maketitle\SIsave@maketitle\let\title\SIsave@title\let\author\SIsave@author
\let\email\SIsave@email\let\affiliation\SIsave@affiliation\let\date\SIsave@date
\makeatother
\title{Supplementary Information for: Finite temperature and pressure structural relaxations including zero-point motion and anharmonicity}

\author{Henricus R.\,A. ten Eikelder$^{1}$}
\email{henricus.ten.eikelder@liu.se}
\author{Florian Knoop$^{1}$}
\author{Florian Trybel$^{1}$}
\email{florian.trybel@liu.se}
\affiliation{1) Department of Physics, Chemistry and Biology (IFM), Link\"oping University, SE-581 83 Link\"oping, Sweden}

\date{\today}

\maketitle
\setcounter{page}{1}
\onecolumngrid
\setcounter{section}{0}

\setcounter{figure}{0}\renewcommand{\thefigure}{S\arabic{figure}}\def\theHfigure{S\arabic{figure}}
\setcounter{table}{0}\renewcommand{\thetable}{S\arabic{table}}\def\theHtable{S\arabic{table}}

\subsection*{S1. The transition with the LaH$_{10}$-120-sample potential}
\begin{figure}[H]
    \centering
    \includegraphics[width=0.85\linewidth]{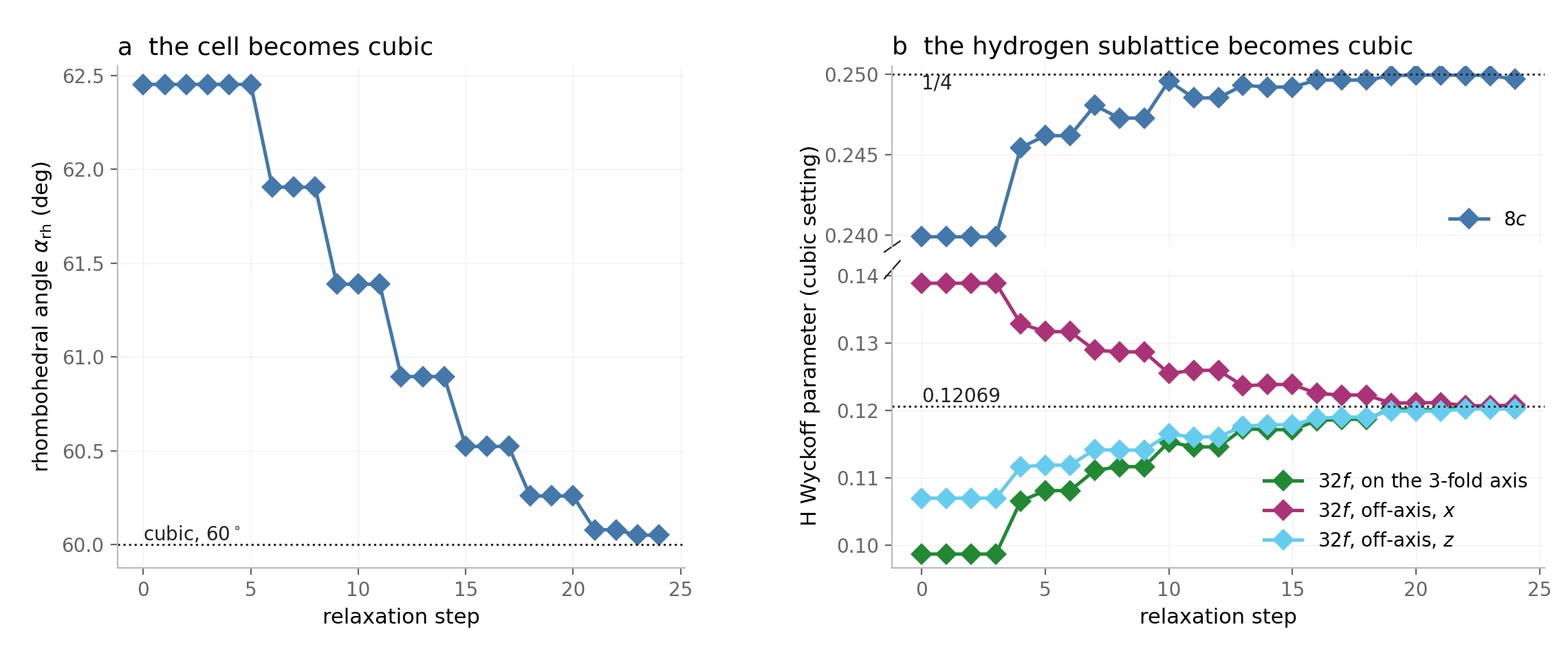}
    \caption{\justifying LaH$_{10}$, 150~GPa, 300~K, $R\bar{3}m\to Fm\bar{3}m$ with the fine-tuned UPET potential from 120 training supercells, in the layout of Fig.~\ref{fig:lah_transition}: (a) rhombohedral angle and (b) hydrogen Wyckoff parameters in the cubic setting against relaxation step. The training configurations are taken from the first three lattice steps of the DFT-force run of Fig.~\ref{fig:lah_transition} and do not reach the cubic endpoint; the relaxation runs from $62.452^\circ$ to $60.050^\circ$, $0.050^\circ$ above the cubic metric, in seven lattice steps, with the $8c$ site at $\tfrac14$ and the three $32f$ parameters merged to 0.1204--0.1208.}
\label{fig:si_lah_mlip}
\end{figure}

\subsection*{S2. Brucite: static cells and the finite-temperature structures}
Table~\ref{tab:si_brucite_static} gives the static PBE$+$D3 cell and that of the fine-tuned potential, each from a variable-cell relaxation of the five-atom cell. The potential reproduces the PBE$+$D3 cell to 0.3\% or better in every quantity listed. The static cell already carries the offsets of the main text: against experiment at 15~K, $c$ is 2.4\% and the interlayer 4.9\% short.

Table~\ref{tab:brucite} lists the final structure of every relaxation of the two series plotted in Fig.~\ref{fig:brucite}, with the deuterium and with the protium mass on the hydrogen sites, together with the single-site neutron O--D interpolated to the same temperatures. Each entry is the last lattice step of its relaxation.

\paragraph*{Isotope mass.} The measured crystal is Mg(OD)$_2$, and the series of Fig.~\ref{fig:brucite} is sampled with the deuterium mass, 2.0141~u, on the hydrogen sites; a second series with the protium mass is plotted behind it. The potential energy surface is the same for both isotopes, and the mass enters only through the thermal and zero-point amplitudes of the sampling and through the phonon frequencies (Section~S5). The two series also differ in the compliance of the lattice update: the deuterium series uses the measured elastic constants \cite{SI:jiang2006brucite} and starts at every temperature from the final cell of the protium series, which used the clamped-ion compliance of the fitted force constants (Section~S3). At the same temperature the two series agree in $a$ to 0.0005~\AA, in the octahedral layer to 0.0015~\AA\ and in $c$ to 0.009~\AA, with differences of either sign, which is within what a single relaxation resolves in $c$. The partition of the expansion is the same: between 30 and 300~K the octahedral layer carries 20\% of the increase of $c$ with deuterium and 17\% with protium, against 79\% in experiment. Over the same range the O--D distance decreases by 0.0026~\AA\ with deuterium and by 0.0015~\AA\ with protium, against 0.0067~\AA\ measured.

\begin{table}[H]
\centering\small
\setlength{\tabcolsep}{4pt}
\begin{tabular}{lccccccc}
\toprule
 & $a$ (\AA) & $c$ (\AA) & oct.\ layer (\AA) & interlayer (\AA) & Mg--O (\AA) & O--D (\AA) & $V$ (\AA$^3$/atom) \\
\midrule
DFT PBE$+$D3 & 3.1684 & 4.6120 & 2.0839 & 2.5281 & 2.1052 & 0.9713 & 8.019 \\
potential PBE$+$D3 & 3.1676 & 4.6185 & 2.0897 & 2.5288 & 2.1063 & 0.9718 & 8.027 \\
neutron, 15~K \cite{SI:chakoumakos1997brucite} & 3.1408 & 4.7270 & 2.067 & 2.659 & 2.0874 & 0.9563 & 8.077 \\
\bottomrule
\end{tabular}
\caption{\justifying \textbf{Brucite static cells.} Variable-cell relaxations of the five-atom cell at ambient pressure with PBE$+$D3 and with the fine-tuned potential. The octahedral-layer thickness is $2|z_{\mathrm O}-z_{\mathrm{Mg}}|\,c$ and the interlayer thickness is $c$ less that. The neutron O--D is that of the single-site model, the like-for-like comparison with the computed distance.}
\label{tab:si_brucite_static}
\end{table}

\begin{table}[H]
\centering\footnotesize
\setlength{\tabcolsep}{4pt}
\begin{tabular}{rccccccccc}
\toprule
$T$ (K) & $a$ (\AA) & $c$ (\AA) & oct.\ layer (\AA) & interlayer (\AA) & Mg--O (\AA) & O$\cdots$O (\AA) & O--D (\AA) & neutron O--D, single-site (\AA) & $\Delta$(O--D) (\AA) \\
\midrule
\multicolumn{10}{l}{\emph{deuterium mass}} \\
15 & 3.1811 & 4.6723 & 2.1058 & 2.5664 & 2.1170 & 3.1559 & 0.9685 & 0.9563 & $+0.0122$ \\
30 & 3.1814 & 4.6668 & 2.1049 & 2.5619 & 2.1170 & 3.1523 & 0.9688 & 0.9566 & $+0.0122$ \\
60 & 3.1814 & 4.6691 & 2.1051 & 2.5640 & 2.1170 & 3.1540 & 0.9686 & 0.9582 & $+0.0103$ \\
90 & 3.1814 & 4.6724 & 2.1057 & 2.5668 & 2.1171 & 3.1563 & 0.9687 & 0.9598 & $+0.0089$ \\
120 & 3.1819 & 4.6736 & 2.1058 & 2.5678 & 2.1174 & 3.1573 & 0.9686 & 0.9590 & $+0.0096$ \\
150 & 3.1825 & 4.6792 & 2.1066 & 2.5726 & 2.1179 & 3.1614 & 0.9684 & 0.9576 & $+0.0108$ \\
180 & 3.1833 & 4.6815 & 2.1071 & 2.5744 & 2.1185 & 3.1631 & 0.9678 & 0.9562 & $+0.0116$ \\
210 & 3.1846 & 4.6801 & 2.1071 & 2.5730 & 2.1191 & 3.1624 & 0.9675 & 0.9546 & $+0.0129$ \\
240 & 3.1858 & 4.6825 & 2.1076 & 2.5749 & 2.1198 & 3.1643 & 0.9671 & 0.9530 & $+0.0141$ \\
270 & 3.1868 & 4.6906 & 2.1091 & 2.5816 & 2.1207 & 3.1701 & 0.9663 & 0.9510 & $+0.0152$ \\
300 & 3.1882 & 4.6841 & 2.1083 & 2.5757 & 2.1212 & 3.1659 & 0.9661 & 0.9499 & $+0.0163$ \\
\midrule
\multicolumn{10}{l}{\emph{protium mass}} \\
15 & 3.1814 & 4.6729 & 2.1062 & 2.5667 & 2.1173 & 3.1562 & 0.9688 & 0.9563 & $+0.0126$ \\
30 & 3.1816 & 4.6700 & 2.1057 & 2.5643 & 2.1172 & 3.1544 & 0.9690 & 0.9566 & $+0.0124$ \\
60 & 3.1818 & 4.6613 & 2.1042 & 2.5571 & 2.1170 & 3.1485 & 0.9692 & 0.9582 & $+0.0110$ \\
90 & 3.1817 & 4.6715 & 2.1057 & 2.5658 & 2.1173 & 3.1556 & 0.9689 & 0.9598 & $+0.0091$ \\
120 & 3.1823 & 4.6730 & 2.1058 & 2.5672 & 2.1176 & 3.1569 & 0.9689 & 0.9590 & $+0.0099$ \\
150 & 3.1830 & 4.6743 & 2.1059 & 2.5684 & 2.1180 & 3.1581 & 0.9686 & 0.9576 & $+0.0111$ \\
180 & 3.1836 & 4.6819 & 2.1074 & 2.5746 & 2.1186 & 3.1633 & 0.9684 & 0.9562 & $+0.0122$ \\
210 & 3.1846 & 4.6802 & 2.1071 & 2.5731 & 2.1191 & 3.1625 & 0.9684 & 0.9546 & $+0.0138$ \\
240 & 3.1858 & 4.6825 & 2.1078 & 2.5748 & 2.1198 & 3.1642 & 0.9682 & 0.9530 & $+0.0152$ \\
270 & 3.1870 & 4.6875 & 2.1087 & 2.5788 & 2.1207 & 3.1680 & 0.9678 & 0.9510 & $+0.0167$ \\
300 & 3.1886 & 4.6932 & 2.1097 & 2.5835 & 2.1217 & 3.1723 & 0.9675 & 0.9499 & $+0.0176$ \\
\bottomrule
\end{tabular}
\caption{\justifying \textbf{Brucite final structures by temperature.} The final structure of every relaxation of the two PBE$+$D3 series plotted in Fig.~\ref{fig:brucite}, with the deuterium mass (upper block, the main series) and with the protium mass (lower block): lattice parameters, octahedral-layer and interlayer thickness, the Mg--O bond, the O$\cdots$O distance across the interlayer gap and the O--D distance between mean positions, with the single-site O--D of the neutron diffraction data of \citet{SI:chakoumakos1997brucite} interpolated to the same temperature and $\Delta$(O--D), the computed O--D minus the neutron one.}
\label{tab:brucite}
\end{table}

\subsection*{S3. Computational details}
\paragraph*{Settings common to all systems.}
Density-functional theory \cite{SI:inhomogeneousEG_HohenbergKohn,SI:Kohn-Sham_selfconsistent} was applied with the projector augmented-wave method \cite{SI:PAW} as implemented in \textsc{Quantum ESPRESSO} \cite{SI:Giannozzi2009,SI:Giannozzi2017,SI:Giannozzi2020}, with PAW datasets from the \textsc{pslibrary} tables \cite{SI:DalCorso2014} and PBE \cite{SI:Perdew_Burke_Ernzerhof} and, for brucite, PBE with Grimme D3 dispersion \cite{SI:grimme2010d3}. Starting structures were relaxed to $10^{-4}$~eV~\AA$^{-1}$ and 0.05~GPa. The position-update guards are a maximum displacement of 2~\AA{} and a minimum interatomic distance of 0.3~\AA. Phonon sums---$F_{\mathrm{ph}}$, $\sigma_{\mathrm{ph}}$ and the densities of state---were evaluated on a $\mathbf{q}$-point grid of the unit cell, the densest of $22^3$, $20^3$, $18^3$, $16^3$ and $14^3$ on which the interpolation of the fitted force constants succeeded. The position loop stops when the mean and the maximum $|\Phi^{(1)}|$ fall below 30 and 75~meV\,\AA$^{-1}$ respectively, unless stated otherwise per system; those tolerances are bounded below by the standard error of the ensemble mean of Eq.~\eqref{eq:fc1}, which falls as $N^{-1/2}$ in the number $N$ of configurations, and by whatever part of the calculator's force error is systematic rather than random. Plane-wave cutoffs were 80/640~Ry for Al, 100/700--800~Ry for the hydrides, 110/880~Ry for brucite. Unless stated otherwise, every $k$-mesh quoted per system below is the mesh applied to the sampled supercell, not to the unit cell. TDEP fits used quantum statistics throughout, supercells of $3\times3\times3$ (Al), $2\times1\times1$, $2\times2\times1$, $2\times2\times2$ and $3\times2\times1$ as listed per hydride below together with the 231-atom LaH$_{10}$ cell of the transition, and the 215-atom non-diagonal cell for brucite, with 16--512 configurations per step as stated per system. The noise-referenced gain of the lattice update described in Methods, $\lambda_l=\lambda_0\,z^2/(z^2+z_0^2)$ with $z_0=3$, is used for the final lattice steps of the brucite series and of the LaH$_{10}$ equation of state; earlier steps used $\lambda_l=\lambda_0$.

\paragraph*{Per system.} Cutoffs are given as rc$_2$ and rc$_3$, the second- and third-order force-constant cutoff radii; the frequency quoted per system is the maximum phonon frequency used to generate the initial force constants; and the single pressure quoted per hydride is the reference state point of that entry, not the extent of the campaign.
\emph{Al}: fcc, PBE, scalar-relativistic PAW datasets from pslibrary 1.0.0, 80/640~Ry, a $\Gamma$-centred $2\times2\times2$ $k$-mesh, Gaussian smearing 0.01~Ry, self-consistency to $10^{-6}$~Ry; sTDEP on $3\times3\times3$ supercells with rc$_2$ 6.0 and rc$_3$ 3.5~\AA, initial force constants from a maximum frequency of 11~THz, 16 configurations per position step and 32 accumulated for the lattice step. \emph{Lanthanum hydrides}: PBE, PAW datasets from pslibrary 1.0.0 with the lanthanum semicore states in valence, $\Gamma$-centred $k$-meshes, sTDEP at 300~K with 16 configurations per position step and 32 accumulated for the lattice step, the transition of LaH$_{10}$ with 40. Each stoichiometry was run at 300~K on a 5~GPa grid from 110 to 170~GPa; Fig.~\ref{fig:LaHx_VP} plots the points computed. LaH$_4$: 100/700~Ry, $4\times4\times4$ $k$-mesh, smearing 0.005~Ry, $10^{-6}$~Ry; $3\times2\times1$ supercells of the 20-atom conventional cell (120 atoms), rc$_2$ 4.0 and rc$_3$ 2.5~\AA, 53~THz, 125~GPa, pressure tolerance 0.1~GPa. LaH$_6$: 100/800~Ry, $6\times6\times6$, 0.01~Ry, $10^{-7}$~Ry; $2\times2\times2$ supercells, 4.0 and 2.5~\AA, 43~THz, 125~GPa, 0.005~GPa. LaH$_9$: 100/800~Ry, $6\times6\times6$, 0.01~Ry, $10^{-7}$~Ry; $2\times1\times1$ supercells, 3.5 and 2.0~\AA, 55~THz, 140~GPa, 0.1~GPa. LaH$_{10}$: 100/700~Ry, $4\times4\times4$, 0.01~Ry, $10^{-6}$~Ry; $2\times2\times1$ supercells for the equation of state and a 231-atom supercell (21 formula units) for the $R\bar{3}m\to Fm\bar{3}m$ transition, 4.0 and 2.5~\AA, 63~THz, 150~GPa, 0.005~GPa. \emph{Brucite}: PBE with Grimme D3 dispersion (zero damping), 110/880~Ry, a $\Gamma$-centred $2\times2\times2$ $k$-mesh on the supercell, Gaussian smearing 0.01~Ry, self-consistency to $10^{-6}$~Ry, for the training supercells; the production series with the fine-tuned potential (the epoch-1140 checkpoint of Section~S4) on a non-diagonal supercell of 43 five-atom cells (215 atoms), the rows of its transformation matrix being $(4,2,1)$, $(2,-1,-2)$ and $(1,3,-2)$, with rc$_2$ 5.67 and rc$_3$ 3.5~\AA, initial force constants from a maximum frequency of 118~THz, 512 configurations per position step reused in the lattice step (1024 per lattice step), quantum statistics with the deuterium mass on the hydrogen sites and the measured elastic constants \cite{SI:jiang2006brucite} as the compliance of the lattice update, every temperature started from the final cell of a series with the protium mass and the clamped-ion compliance (Section~S2), $\lambda_w=0.75$, position-loop tolerances 30 and 75~meV\,\AA$^{-1}$, the noise-referenced gain and a pressure gate per component of 0.08~GPa or twice the standard error of the sampled stress, at 15~K and from 30 to 300~K in 30~K steps at ambient pressure. The static cells of Table~\ref{tab:si_brucite_static} are Quantum ESPRESSO variable-cell relaxations of the five-atom cell with an $8\times8\times6$ $k$-mesh and the cutoffs of the series.

\subsection*{S4. The fine-tuned potentials: training data, fitting and held-out errors}
\begin{figure}[H]
    \centering
    \includegraphics[width=0.92\linewidth]{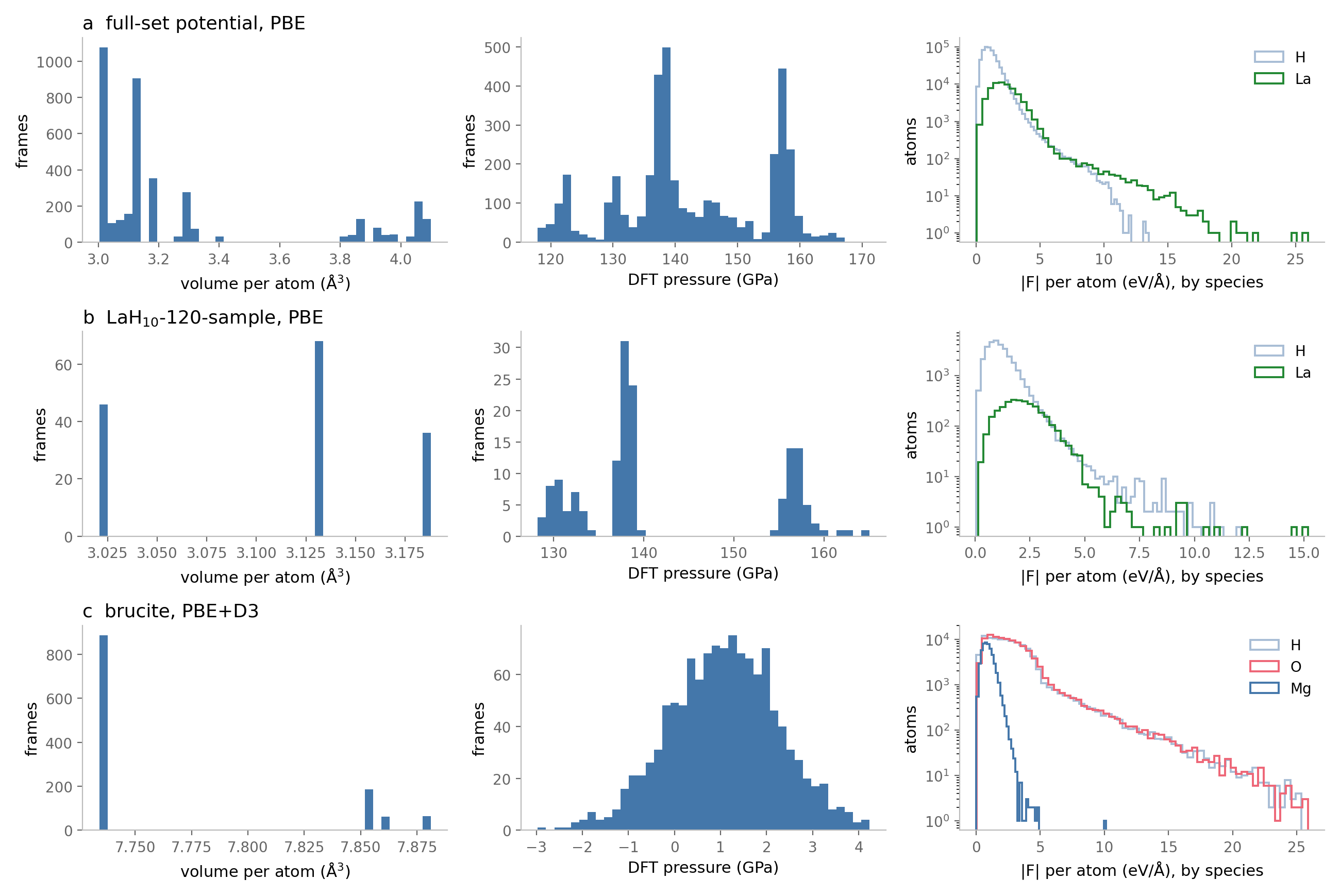}
    \caption{\justifying \textbf{Training-set distributions of the fine-tuned potentials} (train, validation and test frames together). Left: volume per atom; middle: the DFT pressure of the frame; right: force magnitude per atom by species, counts on a logarithmic scale; deuterium is labelled ``H'', the potential carrying hydrogen's chemical identity. Each row is one potential, and the clusters within a row are individual lattice steps of the DFT-force relaxations that produced the frames.}
    \label{fig:si_mlip_data}
\end{figure}

Three fine-tuned potentials are used: the full-set and LaH$_{10}$-120-sample potentials of Sec.~\ref{sec:lah_mlip} and Section~S1, and the PBE$+$D3 brucite potential of Sec.~\ref{sec:brucite}. All share one protocol. The foundation model is pet-omat-l \cite{SI:bigi2026upet,SI:mazitov2025petmad,SI:barrosoluque2024omat24}; every parameter is trained (full fine-tuning, no adapters) with metatrain \cite{SI:metatrain}, on one energy head per DFT setting, with the loss weights 0.1 on the energy, 20 on the forces and 5 on the virial, a 5\% linear warm-up and the learning rates, batch sizes and epoch counts of Table~\ref{tab:si_mlip}; provenance and coverage are in Table~\ref{tab:si_mlip_data}. Reference energies are aligned per DFT setting before training; forces and stresses are used as computed. Every set is split by whole sampling stage, never by frame, so a validation or test configuration always comes from a sampling stage the model has not seen. Every configuration is a thermally displaced supercell generated by a DFT-force relaxation, which is why the distributions of Fig.~\ref{fig:si_mlip_data} are clustered: each cluster is one lattice step of one state point.

\begin{table}[H]
\centering\small
\setlength{\tabcolsep}{3pt}\renewcommand{\arraystretch}{1.15}
\begin{tabularx}{\linewidth}{>{\raggedright\arraybackslash}p{3.2cm}*{3}{>{\raggedright\arraybackslash}X}}
\toprule
 & the full-set potential & LaH$_{10}$-120-sample & brucite PBE$+$D3 \\
\midrule
used for & MLIP points of Fig.~\ref{fig:LaHx_VP}; the transition of Sec.~\ref{sec:lah_mlip} & Fig.~\ref{fig:si_lah_mlip} & brucite series of Sec.~\ref{sec:brucite} \\
source frames & DFT-force relaxations of LaH$_6$, LaH$_9$, LaH$_{10}$ at 300~K & first three lattice steps of one LaH$_{10}$ DFT-force run & DFT-force relaxations, 15 / 150 / 270~K \\
frames train / val / test & 3\,261 / 303 / 319 & 120 / 15 / 15 & 991 / 109 / 96 \\
sampling stages & 203 & 41 & 24 \\
cell sizes (atoms) & 44--231 & 231 & 215 \\
volume per atom (\AA$^3$) & 3.00--4.10 & 3.02--3.19 & 7.73--7.88 \\
DFT pressure (GPa) & 118 to 171 & 128 to 165 & $-3$ to $+4$ \\
force RMS per species (eV\,\AA$^{-1}$) & La 2.73, H 1.44 & La 2.52, H 1.45 & Mg 1.02, O 3.57, D 3.52 \\
\bottomrule
\end{tabularx}
\caption{\justifying \textbf{Provenance and coverage of every fine-tuned potential used.} Train, validation and test frames are split by whole sampling stage. ``DFT-force relaxations'' are \warp{} runs with DFT forces; every frame is a thermally displaced supercell of one of their sampling stages.}
\label{tab:si_mlip_data}
\end{table}

\begin{table}[H]
\centering\small
\setlength{\tabcolsep}{3pt}\renewcommand{\arraystretch}{1.15}
\begin{tabularx}{\linewidth}{>{\raggedright\arraybackslash}p{3.2cm}*{3}{>{\raggedright\arraybackslash}X}}
\toprule
 & the full-set potential & LaH$_{10}$-120-sample & brucite PBE$+$D3 \\
\midrule
learning rate; batch & $1.6\times10^{-4}$; 2 & $5\times10^{-5}$; 1 & $10^{-4}$; 1 \\
epochs run / checkpoint used & 399 / 399 & 1\,990 / 1\,990 & 3\,205 / 1\,140 \\
validation force RMSE at the checkpoint (meV\,\AA$^{-1}$) & 25.1 & 35.3 & 24.9 \\
training force RMSE at the checkpoint (meV\,\AA$^{-1}$) & 12.1 & 4.8 & 17.9 \\
validation energy RMSE (meV/atom) & 1.1 & 0.2 & 1.2 \\
held-out test force RMSE (meV\,\AA$^{-1}$) & 25.8 & 34.5 & 12.3 \\
held-out test energy RMSE (meV/atom) & 1.46 & 0.24 & 0.10 \\
fine-tuning cost (GPU-h) & $\approx$880 & 47 & 620 \\
acceptance test & equation of state and $R\bar{3}m\to Fm\bar{3}m$ endpoint vs DFT & endpoint vs DFT (Fig.~\ref{fig:si_lah_mlip}) & static cell (Table~\ref{tab:si_brucite_static}) \\
\bottomrule
\end{tabularx}
\caption{\justifying \textbf{Fitting and errors.} Validation and training errors are the per-epoch values of metatrain at the checkpoint that was used; held-out test errors are a separate evaluation of that checkpoint on the test split with the calculator the relaxations use. Held-out errors are reported, not used as the acceptance criterion: the test in the last row is structural.}
\label{tab:si_mlip}
\end{table}

\begin{figure}[H]
    \centering
    \includegraphics[width=0.98\linewidth]{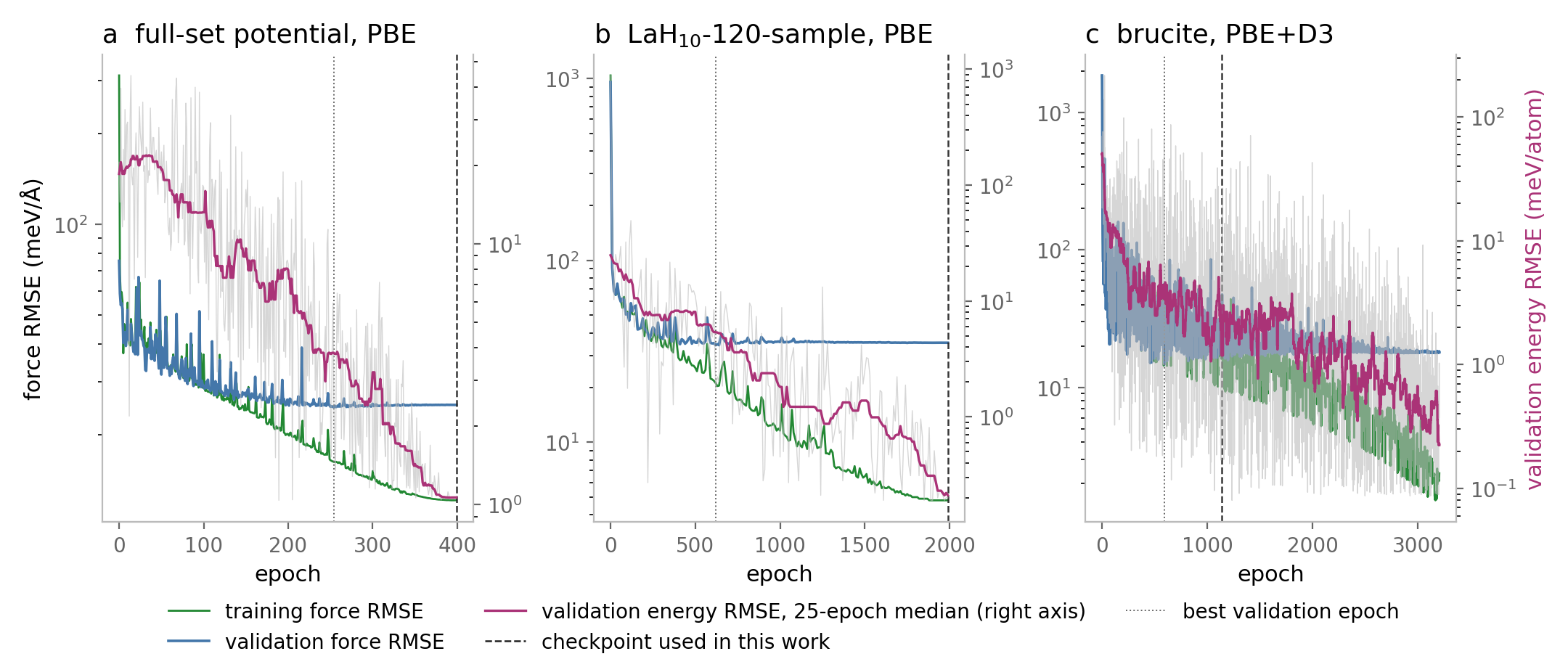}
    \caption{\justifying \textbf{Learning curves.} Training (green) and validation (blue) force RMSE per epoch, the validation energy RMSE per epoch (grey) and as a 25-epoch running median (magenta, right axis), the checkpoint used for the relaxations (dashed; for brucite PBE$+$D3 the epoch-1140 checkpoint of Table~\ref{tab:si_mlip}) and the best validation epoch (dotted).}
    \label{fig:si_mlip_curves}
\end{figure}

\begin{figure}[H]
    \centering
    \includegraphics[width=0.98\linewidth]{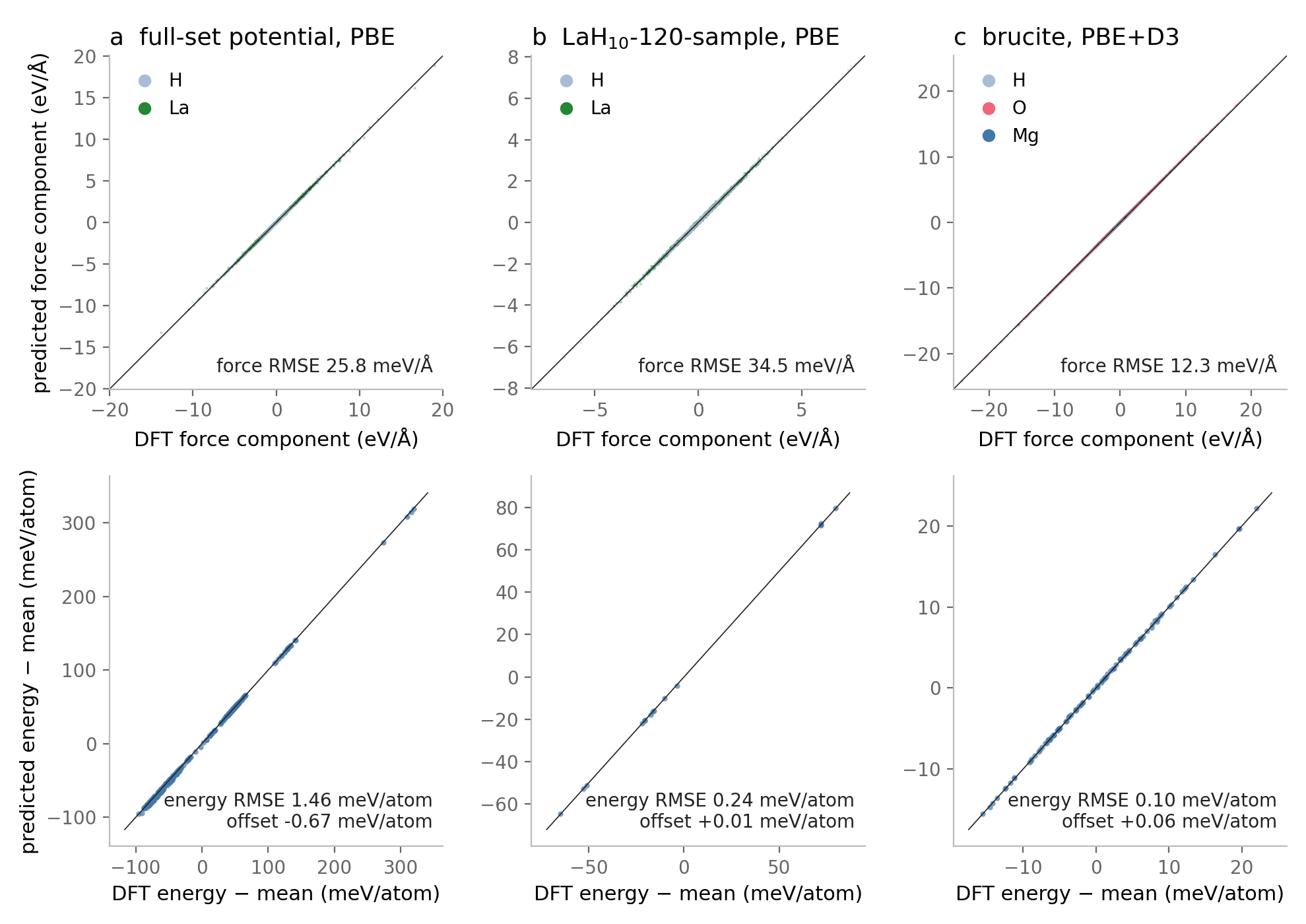}
    \caption{\justifying \textbf{Held-out test sets.} Top: predicted against reference force components, coloured by species (the deuterium sites of brucite are labelled H, as in the potentials' data); bottom: predicted against reference energy per atom, both relative to the test-set mean. Each panel is the checkpoint used for the relaxations, evaluated with the same calculator the relaxations use; the RMSE in the title is the number in Table~\ref{tab:si_mlip}. Oxygen is red, hydrogen (deuterium) light blue, magnesium blue and lanthanum green, as in Fig.~\ref{fig:si_mlip_data}.}
    \label{fig:si_mlip_parity}
\end{figure}

The full-set potential reaches its validation plateau of 25~meV\,\AA$^{-1}$ by epoch 250 (Fig.~\ref{fig:si_mlip_curves}) while the training error keeps falling to 12: the validation stages are different stages of the same relaxations, and the gap is the price of a set that is clustered by construction. The LaH$_{10}$-120-sample potential is that effect taken to its limit, one run's first three lattice steps, and its training error of 5 against a validation error of 35 is expected; Section~S1 shows that the relaxation with it still reaches the DFT endpoint, which is the test that matters. The PBE$+$D3 brucite series uses the epoch-1140 checkpoint, with a validation error of 24.9~meV\,\AA$^{-1}$; the fit continued to epoch 3205, where the validation error settles at 18~meV\,\AA$^{-1}$ against a training error of 3. For this set the validation and test splits are not equivalent: the validation stages are the harder ones, and the same checkpoints score 12.3 (epoch 1140) and 2.1 (epoch 3205) on the test split (Fig.~\ref{fig:si_mlip_parity}). This is the honest reading of a split by sampling stage rather than by frame: the held-out error depends on which stages were held out, which is why the acceptance tests of the last row are structural.

\subsection*{S5. Phonon dispersions: DFT against the fine-tuned potentials}
\begin{figure}[H]
    \centering
    \includegraphics[width=0.92\linewidth]{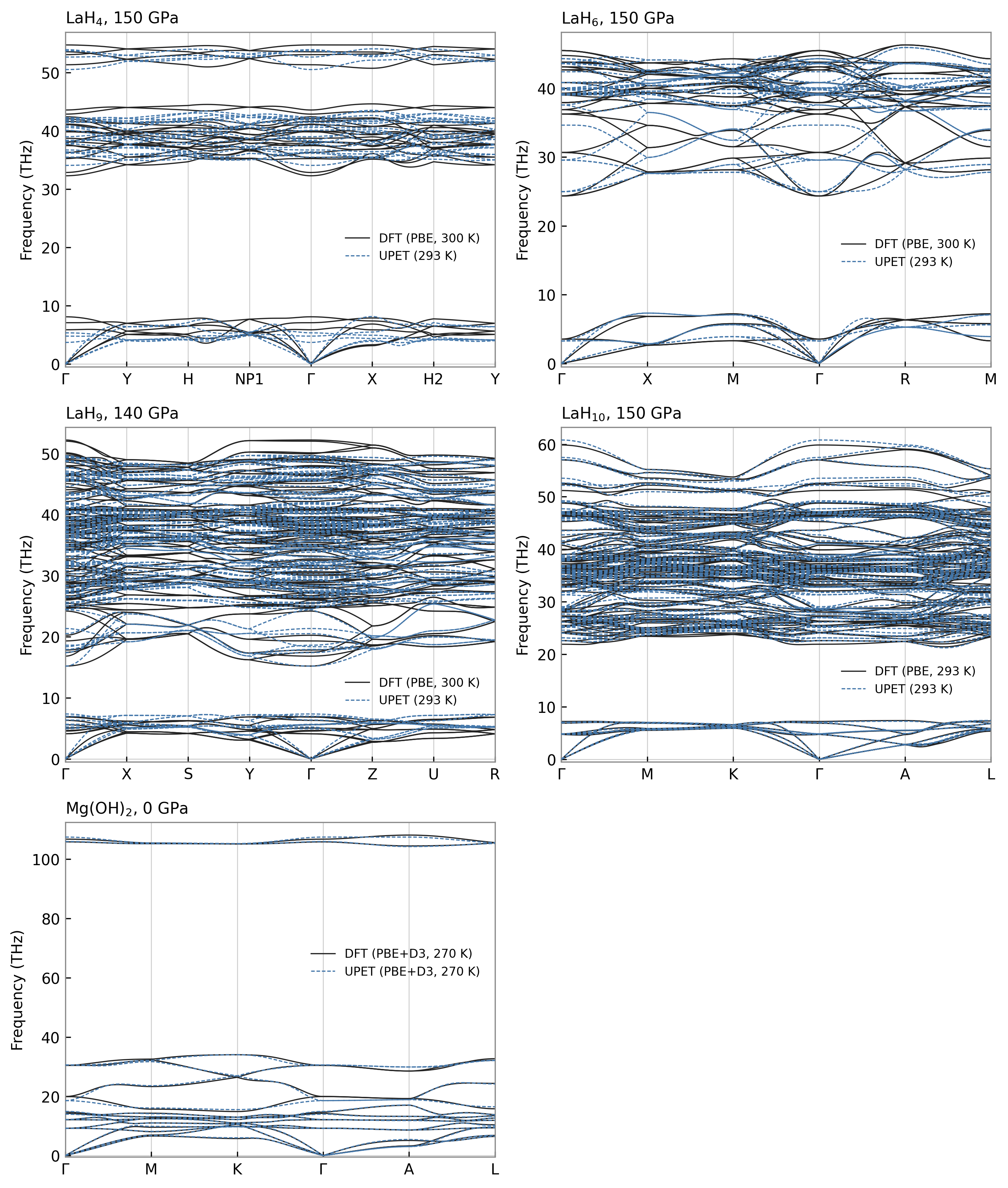}
    \caption{\justifying \textbf{Effective phonon dispersions, DFT against the fine-tuned potentials.} DFT-force relaxations (black, solid) and relaxations with the fine-tuned potential (blue, dashed) of LaH$_4$, LaH$_6$ and LaH$_{10}$ at 150~GPa, LaH$_9$ at 140~GPa and brucite at 0~GPa, at the temperatures in the legends, on the high-symmetry path of each cell; the two paths of a pair are mapped segment by segment onto common tick positions because the two relaxed cells differ slightly.  Brucite is shown with the protium mass in both runs, hence Mg(OH)$_2$.}
    \label{fig:si_dft_upet_ph}
\end{figure}

\begin{figure}[H]
    \centering
    \includegraphics[width=0.55\linewidth]{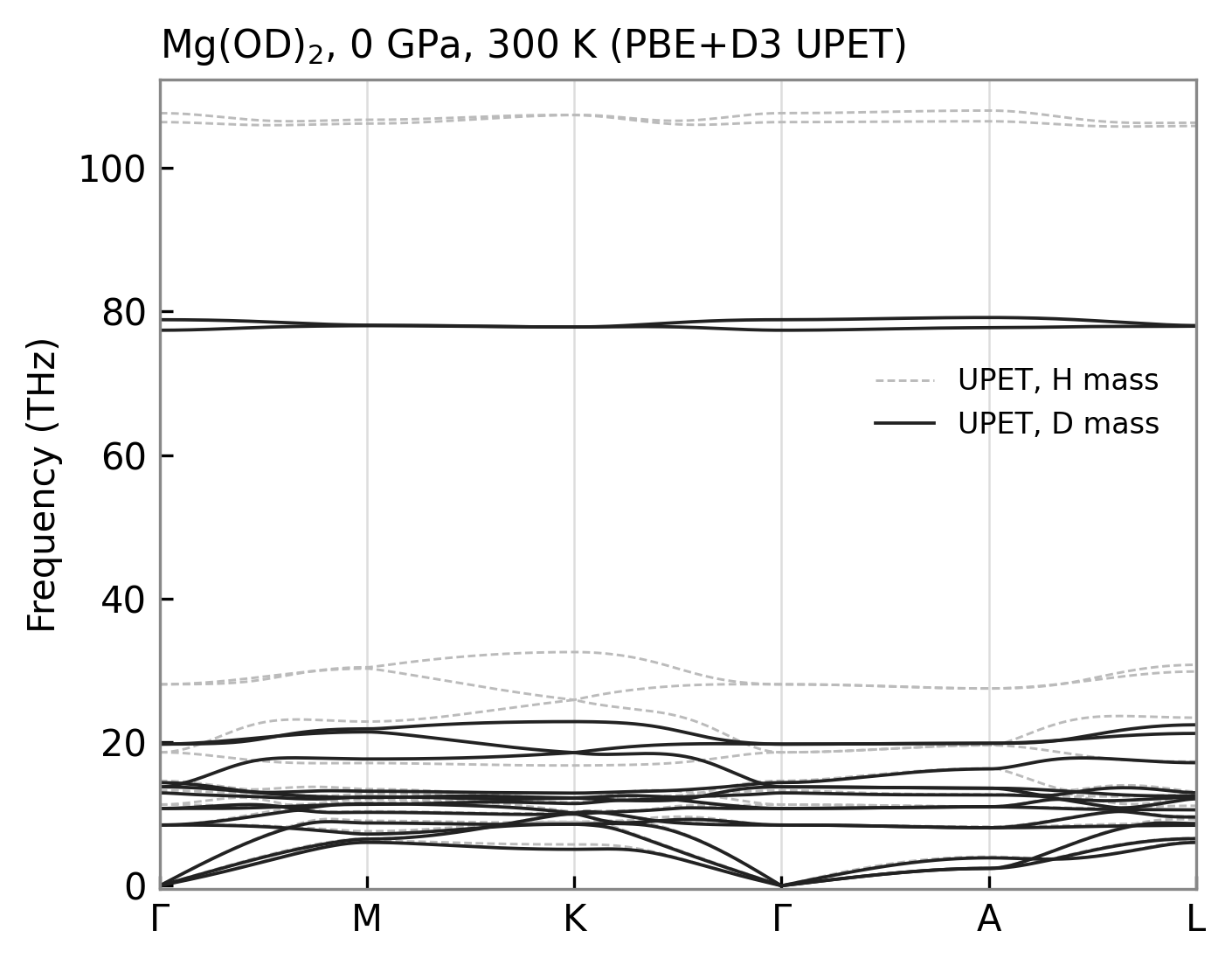}
    \caption{\justifying \textbf{Brucite with the deuterium mass.} Effective phonon dispersion of Mg(OD)$_2$ at 300~K and 0~GPa from the PBE$+$D3 potential with the deuterium mass on the hydrogen sites (black), the last lattice step of the relaxation at 300~K; the same potential with the protium mass at 300~K in grey for comparison.}
    \label{fig:si_brucite_D}
\end{figure}

Figure~\ref{fig:si_dft_upet_ph} compares the effective phonon dispersions of the DFT-force and the potential-driven \warp{} relaxations of the same material and state point. For each pair the DFT and potential steps with the smallest root-mean-square difference of the sorted bands are shown; the force and energy errors of the potentials themselves are those of Table~\ref{tab:si_mlip} and Fig.~\ref{fig:si_mlip_parity}. The root-mean-square differences are 0.92, 0.76, 0.63 and 0.33~THz for LaH$_4$, LaH$_6$, LaH$_9$ and LaH$_{10}$, and 0.36~THz for brucite. LaH$_4$, which is absent from the training set, shows the largest difference, mostly in the hydrogen optical bands. Both brucite runs of this comparison use the protium mass, so the panel compares the two force engines on the same dynamics. The dispersion of Mg(OD)$_2$ with the deuterium mass is shown in Fig.~\ref{fig:si_brucite_D}: the O--D stretching band lies at 78--79~THz, against 106--108~THz with the protium mass, a ratio of 1.36, while the acoustic and lattice modes below 15~THz are nearly unchanged.

\subsection*{S6. LaH$_{10}$: the hydrogen cage across the transition}
\begin{figure}[H]
    \centering
    \includegraphics[width=0.75\linewidth]{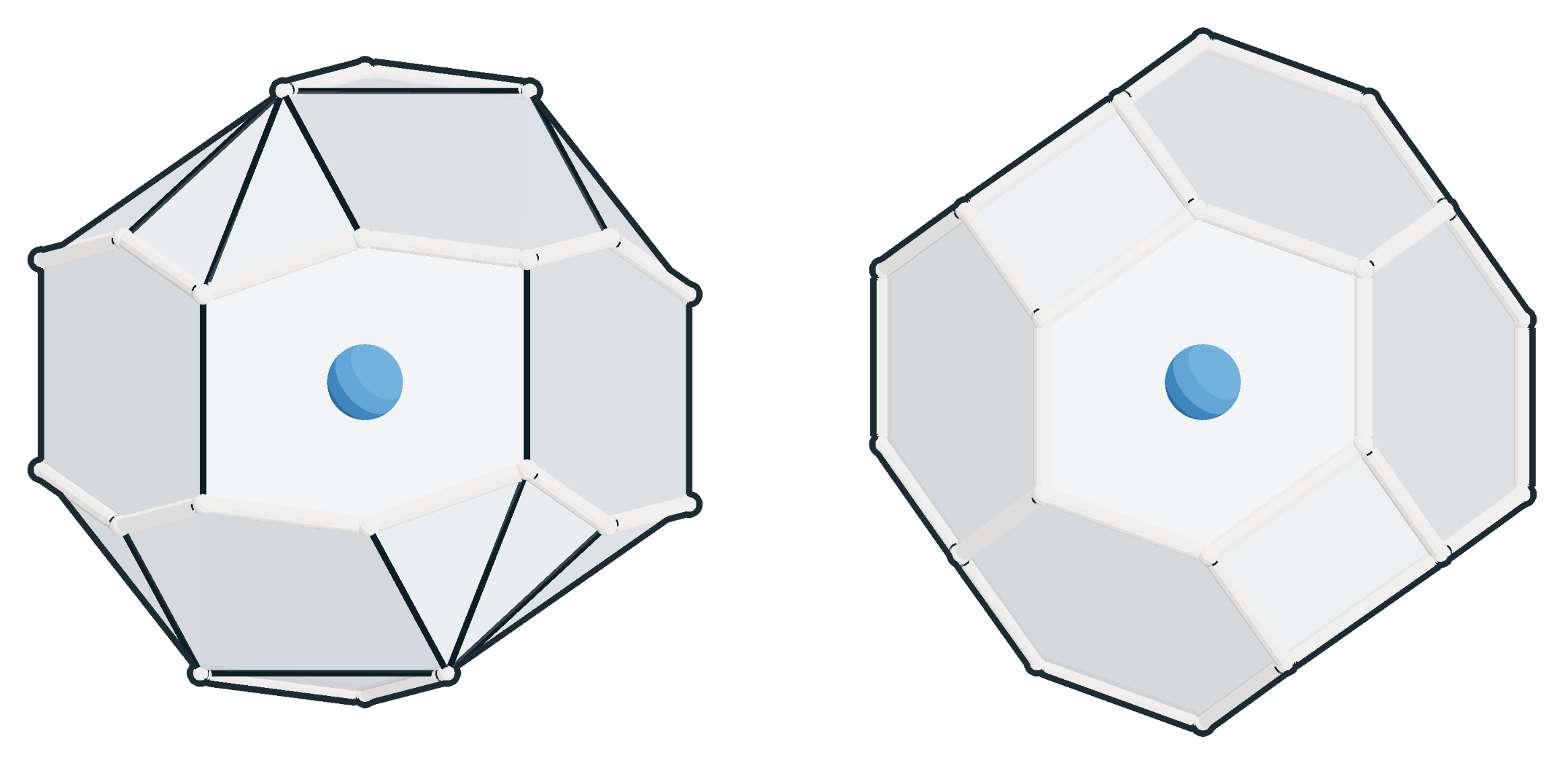}
    \caption{\justifying \textbf{The La-centred H$_{32}$ cage of LaH$_{10}$ at 150~GPa and 300~K.} Left: the rhombohedral ($R\bar{3}m$) static PBE minimum, the start of the relaxation of Fig.~\ref{fig:lah_transition} (rhombohedral angle $62.33^\circ$). Right: the final structure of that relaxation ($59.96^\circ$). Both are viewed perpendicular to the three-fold axis, which is vertical, at the same scale. Lanthanum (blue) at the centre, hydrogen at the vertices; light sticks are H--H distances below 1.30~\AA, the faces are the coordination polyhedron of the lanthanum atom. Rendered with CrysViz.}
    \label{fig:si_lah_cages}
\end{figure}

In $Fm\bar{3}m$ LaH$_{10}$ every lanthanum atom sits in a cage of 32 hydrogen atoms, 8 on the $8c$ and 24 on the $32f$ site, whose faces are six squares and twelve hexagons. Figure~\ref{fig:si_lah_cages} shows this cage at the two ends of the DFT-force relaxation of Fig.~\ref{fig:lah_transition}. In the rhombohedral static minimum the cage is compressed along the three-fold axis, 4.09~\AA\ high against 4.46~\AA\ at the end of the relaxation, the La--H distances spread from 2.05 to 2.27~\AA, and only 30 of the 48 cage edges are short H--H contacts (1.09--1.11~\AA); the others are 1.42~\AA\ or longer, so that the cage closes with triangles at its poles. At the end of the relaxation the La--H distances lie between 2.13 and 2.23~\AA, all 48 edges are H--H contacts of 1.14 to 1.24~\AA, and the cage has the square and hexagonal faces of the cubic structure.

\end{document}